\documentclass[apj]{emulateapj}
\slugcomment{{\sc Accepted for publication in the Astrophysical Journal:} December 21, 2008}
\usepackage{lscape}
\shorttitle{Stellar Population Models and Individual Element Abundances. II.}
\shortauthors{H.-c. Lee et al.}

\begin{document}

\title{Stellar Population Models and Individual Element Abundances II: 
Stellar Spectra and Integrated Light Models}

\author{Hyun-chul Lee, Guy Worthey}
\affil{Department of Physics and Astronomy, Washington State University, 
Pullman, WA 99164-2814}
\email{hclee@wsu.edu, gworthey@wsu.edu}
\author{Aaron Dotter, Brian Chaboyer}
\affil{Department of Physics and Astronomy, Dartmouth College, 6127 Wilder Laboratory, Hanover, NH 03755}
\author{Darko Jevremovi\'c}
\affil{Astronomical Observatory, Volgina 7,
  11160 Belgrade, Serbia and Montenegro}
\affil{Homer L. Dodge Department of Physics and Astronomy, University of Oklahoma, 440 West Brooks, Room 110, Norman, OK 73019-2061}
\author{E. Baron}
\affil{Homer L. Dodge Department of Physics and Astronomy, University of Oklahoma, 440 West Brooks, Room 110, Norman, OK 73019-2061}
\author{Michael M. Briley}
\affil{Department of Physics and Astronomy, University of Wisconsin,
Oshkosh, 800 Algoma Boulevard, Oshkosh, WI 54901}
\author{Jason W. Ferguson}
\affil{Department of Physics,
Wichita State University, Wichita, KS 67260-0032, USA}
\author{Paula Coelho}
\affil{Institut d'Astrophysique, CNRS,
Universit\'e Pierre et Marie Curie, 98 bis Bd Arago, 75014 Paris,
France}
\author{Scott C. Trager}
\affil{Kapteyn Astronomical Institute, University of Groningen, PO Box
800, 9700 AV Groningen, The Netherlands}

\begin{abstract}

The first paper in this series explored the effects of altering the 
chemical mixture of the stellar population on an element by element 
basis on stellar evolutionary tracks and isochrones to the end of the 
red giant branch.  This paper extends the discussion by incorporating 
the fully consistent synthetic stellar spectra with those isochrone 
models in predicting integrated colors, Lick indices, and synthetic spectra.  
Older populations display element ratio effects in their spectra 
at higher amplitude than younger populations. In addition, spectral 
effects in the photospheres of stars tend to dominate over effects 
from isochrone temperatures and lifetimes, but, further, the 
isochrone-based effects that are present tend to fall along the 
age-metallicity degeneracy vector, while the direct stellar spectral 
effects usually show considerable orthogonality.

\end{abstract}


\section{Introduction}

Little is known about the influence of individual element abundances
on the integrated flux of a stellar population.  
As present, it is impossible to derive an accurate age within 10\% from the 
integrated light of a cluster-like single-age and single-abundance 
stellar population, and the primary reason is the complication due to 
abundance ratio effects \citep{wor98}.  The secondary reason is that 
the input ingredients and choices made in even the most sophisticated 
stellar evolution calculations induce scatter in the results 
\citep{char96}.  The tertiary reason is the set of difficulties 
associated with stellar flux knowledge, such as colors, line 
strengths, or spectra, that are needed at each stellar evolutionary 
phase to represent the component stars.  Additional uncertainties 
exist, such as dust extinction, stellar rotation and activity, the 
blue straggler frequency, and the effects of close binaries.

Paper I in this series, \citet{dot07b}, explored the effects of 12 chemical
mixtures on the $L$, $T_{\rm eff}$, and lifetimes of stars along stellar
evolutionary isochrones and upon the opacities needed to calculate the stellar
models. The mixtures explored were solar, $\alpha$-element enhanced, and ten
cases where only one element at a time was enhanced, for elements C, N, O, Ne,
Mg, Si, S, Ca, Ti, and Fe. The mixtures were re-scaled so that the mass
fraction of heavy elements, $Z$, was constant. \footnote{Added to these 12
mixtures were three at non-constant $Z$ and variable $X$ and $Y$ but
constant [Fe/H] with 0.2 dex more Carbon, 0.3 dex more Nitrogen and 0.3 dex
more Oxygen (see \citealt{dot07b}).}

The conclusions from that paper could be summarized by splitting the
elements that were investigated into three categories. ``Displacers''
include O, C, N, and Ne. These are elements that are abundant but
supply less opacity per unit mass than heavier elements. At fixed $Z$,
boosting a displacer element will therefore decrease the opacity. This
leads to shorter stellar lifetimes and hotter stars. ``Boosters''
include Mg and Si. These elements are good opacity sources, so
that boosting a booster will increase opacity and make cooler stars
that live somewhat longer. ``Oddball'' elements defy these trends, and
include Ca and Ti, both of which make cooler stars, that,
nevertheless, have shorter lifetimes. Another oddball is S, which
decreases low-temperature opacity but increases high-temperature
opacity so that its ultimate effect on temperatures and lifetimes is
small. An element worth special mention is Fe, because its
high-temperature opacity has considerable structure. One of its
opacity peaks corresponds to the temperatures characteristic of the
outer edge of the convective core that develops in stars slightly more
massive than the sun. Increasing the Fe abundance therefore strongly
couples to the onset of convective core overshooting effects, and
leads to a strong luminosity effect in the region of the main sequence
turnoff. The temperature, luminosity, and lifetime effects are
illustrated in Paper I, but further consequences will
be illuminated in this paper.

Besides isochrones, the other major ingredient in any population
synthesis effort is some representation of stellar flux for each star
in the isochrone. The effects of individual element abundances on the
spectra of stars have been studied for years, and enormous progress
has been made in finding the structure of stellar atmospheres and
calculating the emergent flux. The two most-cited programs for constructing
model atmospheres of stars are ATLAS \citep{kur70} and MARCS \citep{gus75} and 
we use results from these code in this work.  For spectral synthesis of the 
emergent flux we use SYNTHE \citep{kur70}, 
FANTOM \citep{coe05,cay91}, and SSG \citep{bell94}. For
ordinary population synthesis, empirical spectra or colors can be
used \citep{vaz99}, but for investigating 
element-by-element effects, using synthetic spectra is clearly the way 
forward. Investigations into the effects of individual elements on 
stellar spectra, but with the intent of applying the results to galaxy 
spectra, include \citet{trip95} and \citet{korn05}, who gauged the 
effects of ten individual elements on the Lick system of 25 
pseudo-equivalent width indices \citep{wor94b,wo97}. \citet{serv05} 
investigated 24 elements in spectra that ran from 3500 \AA\ to 9000 \AA\ 
with velocity smoothing appropriate for dynamically hot 
systems such as elliptical galaxies. 

The present paper combines the new isochrones described in Paper I
with greatly extended and updated synthetic spectra in order to obtain 
{\em ab initio} population synthesis models as a function
of {\em individual} element abundances\footnote{\citet{coe07} did a similar 
investigation with a flat enhancement of all the $\alpha$-elements 
over a range of metallicity.}.  All parts of the models, including
high temperature and low temperature opacities, stellar evolutionary
models, and synthetic spectra, include the altered abundances in the
same way.  At present the grid includes only solar $Z$, but
calculations are underway to extend to many different abundances.  This
paper describes the new spectra and their color index results in $\S$ 2, 
Lick index results in $\S$ 3, concluding with a discussion and summary section.

\section{Description of New Model Ingredients}

The new isochrones were described in Paper I \citep{dot07b}, and we 
refer the reader to that paper for details. Twelve chemical mixtures 
were explored self-consistently with both high-temperature and 
low-temperature opacities adjusted properly, and evolution is complete 
only to the end of the red giant branch.  It should be emphasized 
that the models are thus {\em incomplete and should not be used 
blindly when comparing to real stellar populations} until the 
helium-burning phases are properly incorporated\footnote{The contribution 
from the later stellar evolutionary phases to the spectral indices  
can be inferred from Figure 8 in Coelho et al. (2007).}.  This is 
planned for the ongoing improvement of these models. In careful, 
differential ways, the present models give us important pointers to 
the behavior of stellar populations with variable element mixtures, 
but the absolute values of colors and indices are not to be 
trusted.  The mixtures explored were solar, $\alpha$-element enhanced, 
and ten cases where only one element at a time was enhanced (C, N, O, 
Ne, Mg, Si, S, Ca, Ti, and Fe).  The mixtures were re-scaled so that 
the mass fraction of heavy elements, $Z$, was held constant.

\subsection{Synthetic Spectra}

New for the present work is a collection of synthetic spectra from
three different sources. Spectra for the coolest stars, at 3000 and
3190 K for log $g$ = 0, were calculated using the FANTOM synthesis
code and \citet{plez92} model atmospheres.  At 3500 and 3750 K for log
$g$ = 0 and 5 spectra were calculated using the FANTOM synthesis code
and ATLAS model atmospheres exactly as described in \citet{coe05},
from 3000 \AA\ to 10000 \AA\ in 0.02-\AA\ steps.  Spectra for the
medium-temperature stars were calculated using the SSG synthesis code
and MARCS model atmospheres from 3000 \AA\ to 10000 \AA\ in 0.01-\AA\
steps.  There were 9 temperatures in this range, from 4000 K to 6000
K, in 250 K steps. Gravities of log $g$ = 4.5 and 2.5 were used for
all nine temperatures, but log $g$ = 0.5 was included for the three
coolest. Spectra for hot stars, at 7000 K, 8000 K, 10000 K, and 20000
K, were calculated using the SYNTHE synthesis code starting from ATLAS
model atmospheres in high resolution with logarithmic wavelength
spacings. Gravities of $\log g$ = 4.5 and 2.5 were used except the
20000 K models, where 4.5 and 3.0 were used.  
The lines lists were not completely homogeneous between the three
regimes, although they share much in common.  The hot stars were
computed with R. Kurucz (kurucz.harvard.edu) line lists, the 
medium-temperature stars included custom modifications by R. Bell, 
M. Tripicco, and M. Houdashelt, and the cool stars benefitted from 
the TiO re-scalings described in \cite{coe05}.

These spectra were then
rebinned to a common wavelength range of 3000 \AA\ to 10000 \AA\ and a
common linear wavelength binning of 0.01 \AA\ per flux point for
future high-resolution studies, and then rebinned again to a common
linear wavelength binning of 0.5 \AA\ per flux point for more
convenient use for working with colors and spectral
indices\footnote{We have used synthetic stellar spectra with an
elemental enhancement by 0.3 dex in this study.  As discussed in
Tripicco \& Bell (1995), however, a 0.15 dex enhancement is used for
carbon in order to avoid the carbon star abundance regime where carbon
atoms outnumber oxygen atoms.}.

For each star in the grid of spectra described above, multiple
realizations of the same spectrum were calculated with single-element
abundance adjustments, so that the effects of C, N, O, Ne, Mg, Si, S,
Ca, Ti, and Fe could be individually gauged.  For neon enhancement, we use
the scaled-solar spectra because of the small effect on the
optical stellar spectrum from neon.  Any isochrone-level temperature,
luminosity, or stellar lifetime effects that we depicted in Paper I
from neon are, however, included.  Thus, at solar
metallicity, 35 gravity and temperature combinations $\times$ 10
element adjustments (counting the non-adjusted solar abundance
spectrum) makes a total of 350 spectra.

Another convenience is the definition of $R$, which stands for 
``generic heavy element.'' Usually, one refers to relative abundance 
changes relative to iron, such as [O/Fe] or [Mg/Fe]. However, this 
stops making sense in the case of wanting to change Fe with respect to 
other elements; i.e., using [Fe/Fe] leads to nonsense. So we use the 
``generic heavy element'' R to stand for ``all heavy
elements that remain scaled with the solar mixture, where all other
element tweaks must be specified.''  With that notation, [Fe/R] is 
perfectly acceptable, and 
the specification of constant [R/H] means that all elements except the 
one (or, more generically, ones) under consideration are held in solar 
lockstep.

\subsection{Synthetic Spectra Accuracy}

An interpolation routine was constructed so that spectra of arbitrary
temperature, gravity, and abundance mixture could be produced on
demand. Linear interpolation in the log of the flux seemed to give the
most predictive results. Interpolation was done for all input variables:
temperature, log $g$, and the whole set of abundance parameters. 
Tests in which one star in the grid was
compared with its interpolated version calculated from the flanking
temperature grid points indicate that, within the dense
4000-6000 K part of the grid, interpolations yield better than 1\%
accuracy at any given flux point, since the spectra change more with
temperature than with the other parameters. Given the nonlinear temperature
response of many lines, e.g., \citet{gr01}, this is encouragingly
good. Doubtless, in the future, we will demand a denser temperature
grid, but 1\% accuracy is better than we need for the present study.

Toward the more fundamental question of how well the synthetic spectra 
match real spectra, there is little we can add to discussions in 
\citet{mc07,coe05,korn05,trip95}.  \citet{serv05} 
did discover some too-strong lines due to chromium
(that is, mistakes in the line list) via stellar comparisons, but
these lines do not affect the present conclusions at all.  
More worrisome is silicon 
because the features that show Si effects are mostly SiH lines in the 
blue, but these line lists may be immature (R. Peterson, private 
communication).  All conclusions regarding Si should be regarded with
great suspicion until this question is satisfactorily resolved.

In Figures 1 and 2, we compare the high resolution spectra of the Sun
and Arcturus with the grid-interpolated synthetic spectra near the
regions of the H$\beta$ and the Mg $b$ lines, respectively.  For the
solar parameter we employ T$_{eff}$ = 5777 K, $\log g = 4.44$, and
[Fe/H] = 0, and for Arcturus T$_{eff}$ = 4290 K, $\log g = 1.9$, and
[Fe/H] = $-$0.7 with [$\alpha$/Fe] = +0.4 dex (e.g.,
\citealt{pet93,gri99,car04,ful07,koch08}).  The top panels are the
solar spectra and the bottom panels are Arcturus.  The thicker
(black) lines are the observations and the thinner (red) ones are the
synthetic.  The degraded low-resolution spectra, Gaussian-smoothed to 
8 \AA\ FWHM or about 200 km s$^{-1}$ --- somewhat better than the best 
Lick-system resolution, are
overlaid on top of the high-resolution spectra.  We investigate the
individual element effects over a rather broad wavelength region for
the Lick indices and broadband colors in this paper.  In this context,
it is useful to find that the notable mismatches between the
synthetic and the observed spectra in the high-resolution comparisons
are smeared out and become subtle at 
low resolution\footnote{Due to assumptions about scattering versus
absorption in the SSG synthesis code, the bottoms of the saturated
lines are artificially flattened.  Some more comparisons between our
synthetic spectra and Arcturus can be found at
http://astro.wsu.edu/hclee/NSSPM$\_$II$\_$Arcturus.html}.

\placetable{tab:color1}

Table \ref{tab:color1} presents some comparisons between synthetic 
colors and empirical color behavior for the stars in the temperature range 
4000 to 6000 K.  Columns 2-5 refer to {\em synthetic} spectra, and columns 6-8 
refer to the {\em empirical} calibration of \citet{wl08}.  The synthetic 
spectra were converted to $UBVR_CI_C$ colors as in \citet{wor94a} 
using Bessell's (1990) filter responses zeroed to Vega's colors.  There 
is at least a 10 mmag uncertainty just from that procedure, and 
probably at least 20 mmag for $U-B$. The ``dwarf'' is $\log g = 4.5$, 
and the ``giant'' is $\log g = 2.5$.  Columns 2 and 6 present the 
colors for the star listed, synthetic and empirical, respectively.  
The most serious mismatches are the $U-B$ color for the 6000 K dwarf 
and for the 4000 K and 5000 K giants.  These facts suggest some room 
for improvement in model atmospheres and 
line lists (especially at shorter wavelengths).  

The remaining columns present color changes induced by either an 
abundance increase of 0.3 dex or a temperature increase of 250 K 
(color changes in milli-magnitudes). Column 4 is a complete scaled-solar 
enhancement of 0.3 dex in every element and with new atmospheres 
calculated, while Column 3 is a 0.3-dex enhancement of only the 24 
elements that are explicitly traced in the spectral library, a 
superset of those followed in the isochrone library.\footnote{They are 
C, N, O, Na, Mg, Al, Si, S, Cl, K, Ca, Sc, Ti, V, Cr, Mn, Fe, Co, Ni, 
Cu, Zn, Sr, Ba, Eu.}  Comparison of columns 3 and 4 indicates that, 
for the ``heart" of the spectral library between 4000 and 6000 K, 
the sum of the available one-by-one element tweaks approximately equals 
the scaled-solar analog operation. The list of 24 elements in linear 
combination thus appears to approximate the full-blown calculation of 
a spectrum in most cases.  This allows the approximate calculation of 
a new spectrum at arbitrary composition in an eyeblink rather than 
many minutes for a whole new synthetic spectrum calculation.

In general, the synthetic colors are within a few hundredths of a
magnitude of the empirical colors and track the empirical color
responses to abundance and temperature in an approximate way.  
The worst outlier is the $U-B$ 
color for the 6000 K giant, in which the color gets redder rather than 
bluer with increasing temperature (see also \citealt{mc07}).  Strong 
conclusions are not possible with this comparison, but the agreement 
is fairly encouraging.


\placetable{tab:color2}
\placetable{tab:color3}

Tables \ref{tab:color2} and \ref{tab:color3} show the spectral effects of
element-by-element enhancements of 0.3 dex (except carbon which is increased 
by 0.15 dex, see footnote \#4) on some selected color indices for the stars 
in the temperature range 4000 to 6000 K.  In Table \ref{tab:color2} 
the elements were rescaled to constant heavy-element fraction $Z$, 
while in Table \ref{tab:color3} $Z$ was allowed to increase.  One 
technical subtlety is that, since neon is not tracked in the spectral 
library, the only effects that appear in Table \ref{tab:color2} for Ne 
are due to the re-scaling, and there are no effects at all in Table 
\ref{tab:color3}.  Table 3 can be visually explained by taking a 
look at the spectra\footnote{They are given at 
http://astro.wsu.edu/hclee/ NSSPM$\_$II$\_$Color.html}.  Figure 3 displays 
one example.  Here the flux ratio between 0.3 dex Mg-enhanced spectra and 
that of the solar-scaled are shown for a 4000 K giant star.  The $B - V$ 
becomes bluer (e.g., more $B$-band flux over $V$-band flux) and $V - R$, 
$V - I$ become redder (e.g., comparably more $R$- and $I$-band fluxes 
over $V$-band flux).  The $U - B$ also becomes similarly bluer just like 
$B - V$ because the $B$-band filter response curve goes all the way 
to $\sim$5400 \AA.  Now the effects of 
Mg that we read from Table 3 are clearly understood from this illustration.  
It is confirmed that Mg is the most important $\alpha$-element that 
influence the $U-B$ and $B-V$ colors as already described in \citet{cas04}.  

Comparing corresponding entries in the two tables, especially at the 6000 K 
giant, one sees that oxygen have little effect 
on the spectrum by themselves; it is the effect of the decrease of the rest 
of the elemental abundances that causes the bulk of the spectral 
change in the constant-$Z$ case.  Unsurprisingly, it is also clear from 
the table that cooler stars are more susceptible to element-by-element 
effects than warmer stars, at least at optical wavelengths.

\subsection{Integrated-light Color Results}

\placetable{tab:color4}

Table \ref{tab:color4} presents color results for the isochrone-summed models
assuming a \citet{sal55} initial mass function and using the \citet{wor94a}
machinery for some disparate ages of populations.  The bolometric correction
(BC) row includes both isochrone effects and direct stellar spectral effects 
with the caveat that our synthetic spectra are limited in wavelength coverage, 
so the fluxes at given wavelengths were normalized to the older low-resolution 
\citet{wor94a} flux library at the same $Z$.  Thus, only effects that affect 
wavelengths between 3000 and 10000 \AA\ are considered, and there may be 
additional, small effects present that are not accounted for.  However, in 
absolute value, all of the BC shifts are quite small, the largest being 6\% 
for the case of all $\alpha$-elements enhanced.  It is unlikely that element 
ratio effects will be of major concern for the mass estimation of clusters 
and galaxies from their luminosities (see also \citealt{cas04}).

The other immediate conclusion from Table \ref{tab:color4} is that 
older populations show larger spectral effects due to 
element-by-element abundance changes.  This is a straightforward 
consequence of the earlier conclusion that cooler stars are more 
sensitive; the light of older populations is dominated by stars that 
are cooler than those present in younger ones\footnote{This last 
statement is valid for ages $>$ 1 Gyr.  For younger ages, IR 
fluxes from AGB and TP-AGB stars complicate this status (see also 
\citealt{lee07}).}.  Oxygen and neon tend to make the colors 
bluer, while species that contribute more to the lines in the spectrum 
generally make them redder.  Exceptions can be easily explained if one 
knows where the lines contribute the most.  For instance, magnesium has 
most of its absorption near 5100 \AA, i.e., in the $V$ band, so adding 
Mg makes $B-V$ bluer while it makes $V-R$ and $V-I$ redder.  The former, 
the bluer $B-V$ is also partly due to the increase in the flux in the region 
around 4000 \AA\ from the Mg-enhanced spectra as shown in Figure 3 and 
described in \citet{cas04}.  The latter, the redder $V-R$ and $V-I$ also 
reflect the cooler red giant branch as illustrated in figure 8 of Paper I.  
Figure 4 displays this Mg-enhanced integrated case at 5 Gyr (thick line) 
over the single star case of a 4000 K giant (thin line).  The populations 
respond rather like the stars do, as can be seen from Figure 4.  Based on 
this, one would suspect that isochrone-caused effects are relatively minor 
though non-negligible, and this conclusion will be confirmed and amplified 
in the following section.

\section{Results in Lick Index Diagrams}

The effects of element by element enhancement on the Lick indices
\citep{wor94b} are described in this section\footnote{In this study, we 
mostly describe the elemental effects on the Lick indices from 
the {\it integrated} spectra.  Prompted by the referee's suggestion, 
however, we have looked into those elemental effects on the Lick indices 
at the stellar level.  Some examples can be found at 
http://astro.wsu.edu/hclee/NSSPM$\_$II$\_$Lick.html}.  The synthetic indices 
are not very accurate in absolute predictions \citep[c.f.][]{korn05,serv05} 
so we employ a differential approach in which the fitting functions of
\citet{wor94b} and \citet{wo97} are used as the zero point, and delta-index 
information as a function of element ratio is incorporated via 
measuring the synthetic spectral library.  This procedure is similar
to that of previous investigations (e.g.,
\citealt{t00a,t00b,ps02,tmb03,lw05,sch07}) but more sophisticated since
an entire grid of delta-index information was used. That is, 350 spectra 
at solar $Z$, plus similar grids for 4 other $Z$ values for this work, 
as opposed to 2 or 3 synthetic stars at solar abundance only for previous works.

To maintain an exact correspondence with Paper I, all the elements are 
0.3 dex enhanced, except carbon, which is enhanced 0.2 dex\footnote{ 
As footnote \#4 says, the carbon-enhanced spectra are generated with 0.15 dex 
carbon enhancement.  But the carbon-enhanced isochrones that we presented 
in Paper I are of 0.2 dex carbon enhancement.  In order to be consistent, 
for the stellar population synthesis calculations, we extrapolated those 
0.15 dex carbon-enhanced spectra to 0.2 dex enhancement in order to match 
the carbon-enhanced isochrones.  The figure 1 of Paper I, in fact, needs 
to be corrected.  The filled box for the case of carbon should be located 
near $-$0.04 instead of near 0 in terms of [Fe/H].}.  In the 
cases of carbon-, nitrogen-, and oxygen-enhancement, we investigate 
both at fixed total metallicity and at fixed [Fe/H] (and [R/H]). The
cases at fixed [Fe/H] are denoted with plus sign in the figures (e.g., 
C+, N+, O+).  Also in the figures, solar-scaled solar metallicity 
predictions are connected by a solid line from 1 Gyr to 12 Gyr and the 
element-enhanced cases are marked at 1, 2, 5, and 12 Gyr.

We have selected Fe5406 as a reference index in most 
plots.  Among eight Lick iron indices (Fe4383, Fe4531, 
Fe5015, Fe5270, Fe5335, Fe5406, Fe5709, Fe5782) we predict that Fe5406 
is insensitive to every element except iron (see Figures 10 $-$ 11 and 
Table 5 and online spectra described in footnote \#9), 
making it a convenient independent variable. 

As has already been mentioned in the literature \citep{lw07}, for most 
of cases the isochrone effects are relatively minor compared to the 
stellar spectral effects (see also \citealt{sch07}).  For some 
cases, however, isochrone effects are non-negligible and quite
important (cf.\ H$\beta$ in Figure 11; see also figure 17 in \citealt{coe07}).

It may be worthwhile to mention for clarity that ``isochrone effects''
indicate the temperature, luminosity, and stellar lifetime effects
from element ratio changes, and, in the case of the iron-enhanced
mixture, the altered [Fe/H] value that goes into the empirical fitting
functions in our experiment.  When calculating observables, we would
still use a scaled-solar-ratio spectral library.  These isochrones
were presented in Paper I.  In the present Paper II we add the
detailed spectral effects due to element by element enhancement at the
stellar atmosphere/stellar flux level.  The ``direct stellar spectral
effects'' are the ones that come purely from the emergent spectra with
isochrones held fixed.

[Carbon and Nitrogen: CN$_2$]: Lick indices CN$_1$ and CN$_2$ have 
identical central bandpasses, but CN$_2$ has a narrower blue continuum 
which makes the CN$_2$ somewhat less prone to abundance variations 
other than those due to elements C and N.  We compare CN$_2$ with 
G4300 as carbon, nitrogen, oxygen, silicon and iron are enhanced in 
Figure \ref{fig05}.  The upper panels show that CN$_2$ is both carbon- 
and nitrogen-sensitive.  Nitrogen's effect is more prominent (partly 
because carbon is enhanced only by 0.2 dex compared to the 0.3 dex 
nitrogen enhancement).  The bottom right panel also suggests that 
silicon and iron affect CN$_2$ in a non-negligible way when stellar 
populations become older than 5 Gyr.  We can tell from Table 5 that 
the silicon effect is mostly from the stellar spectra, while the iron 
effect is largely from the isochrones.  The oxygen-enhanced case is 
displayed in the bottom left panel. It shows that simply adding O 
alters CN$_2$ and G4300 by a small amount (increases at young ages and 
decreases at old ages; the latter via consumption of more C into the 
CO molecule via molecular equilibrium balance\footnote{ It is 
worth reiterating what Schiavon (2007) noted in his footnote \#5.  
Carbon-enhancement causes the opposite effect from oxygen-enhancement 
and vice versa.  This is because the high dissociation potential of 
CO molecule.  Therefore, at cooler temperatures, 
more carbon translates to more CO, resulting in less oxygen and vice 
versa.}), but if $Z$ is held constant both indices decrease much more 
due to the displacement of C and N to lower abundance because of the 
O-enhancement and fixed sum.  G4300 is N-insensitive. Also, we note 
that CN$_2$ has little sensitivity to Mg-, S-, Ca-, or Ti-enhancement.

[Carbon: G4300 vs. C$_2$4668]: In Figure \ref{fig06}, G4300 and 
C$_2$4668 are plotted as a function of Fe5406.  The left panels show 
C- and Fe-enhanced models, while the right panels show the effects of 
Si and Ti enhancement.  G4300 and C$_2$4668 are known to be good 
carbon indicators among Lick indices along with CN$_1$ and CN$_2$.  
The left panels show that this is indeed the case.  In the bottom left 
panel, however, it can be seen that C$_2$4668 is also highly iron 
sensitive, as indicated by its former name, Fe4668 
\citep{wor94a}\footnote{We find from Table 5 that the Fe-sensitivity 
of C$_2$4668 (also CN$_2$ and Fe5015) is mostly an isochrone effect.  
However, it is, in fact, not because of the temperature and/or 
luminosity changes, but because of the [Fe/H] changes that go into the 
fitting function that we use for the index calculation.  As one can 
find from Figure 1 of Paper I, [Fe/H] = 0.268 for the 
Fe-enhanced isochrones compared to $-$0.225 for the $\alpha$-enhanced 
ones at constant solar metallicity, $Z$.  Clearly, future fitting-function 
work will need to be more meticulously defined in terms of abundance 
parameters, and not locked to [Fe/H] necessarily.}.  Contrary to C$_2$4668, 
G4300 shows a negligible iron sensitivity.  Furthermore, the right panels 
show that C$_2$4668 is influenced by Si and Ti, whereas G4300 shows 
only slight Ti sensitivity.  Both G4300 and C$_2$4668 show little 
sensitivity by N-, Mg-, S-, or Ca-enhancement.

The prospects for disentangling C, N, and O are now fairly 
good, because it appears that there is a lot of sensitivity among 
various indices. However, at the moment it appears that, with the 
indices available there will be considerable degeneracy among the 
three quantities. In the future, adding NH, CH, and CO features may help 
considerably (e.g., \citealt{yong08,mart08,emq08}).

[Calcium: Ca4227 vs. Ca4455]: Figure \ref{fig07} compares Ca4227 and 
Ca4455 with Fe5406.  Carbon-, nitrogen-, and oxygen-enhanced cases at 
fixed [Fe/H] are displayed at the left panels, while right panels 
depict calcium- and iron-enhanced cases at fixed $Z$.  It is clear from 
upper right panel of Figure 7 that Ca4227 is significantly boosted 
with calcium enhancement, the effect increasing with age.  To a lesser 
degree, Ca4227 is also affected by C, N, O, and Fe.  Carbon- and 
nitrogen-enhancement make Ca4227 weaker by 0.3 \AA, whereas the
oxygen-enhancement does the opposite mostly due to their effects 
at the blue-continuum (see also \citealt{pro05}).  

Contrary to Ca4227, Ca4455 is hardly influenced by any of those elements.  
According to this study, Ca4455 is found to be the most 
element-enhancement-free Lick index.  This is consistent with 
the previous findings by \citet{trip95} and \citet{korn05} although 
our presentation is based on a large spectral grid weighted by isochrones 
rather than 3 stars.  One concern, however, is that two recent data sets 
of Milky Way globular clusters, by \citet{coh98} and \citet{puz02}, both
significantly disagree with theoretical model predictions and also 
each other (Figure 3 of \citealt{lw05}).  According to Tables 1 $-$ 3 
of \citet{trip95}, Ca4455 has the strongest dependence on the bandpass 
placement (wavelength shift error) among Lick indices.  This is because 
the blue-continuum and the index bandpass of Ca4455 share the strong Ca 
absorption line feature near 4455 \AA\ and consequently cancel out 
its effect, but only if the wavelength match is perfect.  This phenomenon 
at the stellar level with 10 km/sec velocity dispersion is displayed 
in Figure 8.  Both Ca4227 and Ca4455 are rather insensitive to 
Mg-, Si-, S-, or Ti-enhancement.

[Titanium: Fe4531 and Fe5015]: Fe4531 and Fe5015 are displayed in 
Figure 9 with Fe5406.  TB95, \citet{t97}, KMT05, LW05, and 
\citet{serv05} indicate that these two indices are titanium-sensitive, 
and we confirm that their titanium sensitivity is indeed strong and 
almost comparable to the iron sensitivity.  Fe5406, on the other hand, 
demonstrates little titanium sensitivity.

[Balmer lines: H$\beta$, H$\gamma_{A}$, H$\gamma_{F}$, H$\delta_{A}$,
H$\delta_{F}$]: Balmer lines are widely used as an age indicator 
because of their nonlinear temperature sensitivity in stars, tracing 
better than many indices the temperature of the main-sequence turnoff.  
However, \citet{lw05} and earlier work \citep{wor94b,tmk04,coe07} 
found that they are also abundance sensitive to some degree.  In 
Figures \ref{fig10}--\ref{fig18}, we look into their element by element 
sensitivity in detail.

Effects of the individual 10 chemical elements' enhancement on the 
H$\beta$ and Fe5406 are shown in Figure 10.  The effects of carbon, 
nitrogen, and oxygen enhancement are of great importance but they are 
relatively difficult to understand here because our experimental setup  
preserves the total metallicity.  In other words, it is not 
straightforward to determine whether we are seeing the effects of C-, 
N-, O-enhancement or whether we are seeing other elements (such as Mg 
and Fe) counter-effects due to their depression.  Hence, in Figure 11, 
we display C-, N-, O-enhancement cases both at the fixed total 
metallicity and at the fixed [Fe/H] (the plus signs).  It is seen from 
left panels of Figure 11 that unlike nitrogen, oxygen at young ages and 
carbon at old ages influence H$\beta$.  The top and bottom right 
panels of Figure 11 illustrates that it is mostly the isochrone-level 
effects of carbon and oxygen enhancement that affect H$\beta$.  
Figure 10 further shows that H$\beta$ is similarly altered by the 
Mg enhancement and in this case it is mostly due to the synthetic 
spectra (see also Table 5 and Figure 12).

In the lower right panel of Figure 10 (also in the bottom panels of 
Figure 11), a grid with a range of metallicity ($\mathrm{[Fe/H]} = 
-2.0$, $-$1.0, $-$0.5, 0.0, and 0.5) is displayed at 1, 2, 5, and 12 
Gyr.  These additional calculations are of {\em solar-scaled} chemical 
mixtures \citep{dot07a}.  We have also depicted $\alpha$-elements 
enhanced cases both at the fixed $Z$ and at the fixed [Fe/H] 
(Alpha+)\footnote{The $\alpha$-elements enhanced case at the fixed [Fe/H] 
(Alpha+) is $[\alpha/\mathrm{Fe}] = +0.4$ dex at $\mathrm{[Fe/H]} = 0$ 
using the \citet{dot07a}.  The [Fe/H] of the $\alpha$-elements enhanced case 
at the fixed $Z$ (Alpha) is $-$0.225 using the Paper I.}.  
It is intriguing to note that the 
effect due to the Fe-enhancement by 0.3 dex is seen over the 
$\mathrm{[Fe/H]} = 0.5$ grid line.  This is because these grids only reflect 
the {\em mere isochrone effects} of iron variation with {\em solar-scaled} 
chemical mixtures, while the 0.3 dex Fe-enhancement case shows both 
isochrone and spectral effects combined. 

We find that the H$\beta$ becomes weaker in the $\alpha$-elements enhanced 
case at fixed [Fe/H] mostly because of the effects of Mg (both isochrone and 
spectral as one could see from Table 5), 
while it stays nearly unchanged in the $\alpha$-elements                 
enhanced case at the fixed $Z$ mostly because of the reflection of
depression of Fe.  At 12 Gyr, a comparison of integrated spectra uncovers 
that adding Mg brings down the blue continuum, making H$\beta$ weaker
\footnote{This was hinted in \citet{trip95}.}, 
but adding Fe brings down the central bandpass as well as the red 
continuum, the net effects of which tend to cancel out effects from Fe.  
Alpha-enhancement at fixed $Z$ mostly reflects the decrease of Fe due to 
dilution rather than overt, direct spectral effects from $\alpha$-elements.  
They are displayed in Figures 12 to 14.  The solid lines are the index bandpass 
and the dashed lines are the blue- and the red-continuum edges.  

Figures 15 and 16, on the contrary, display that both H$\gamma$ and
H$\delta$ become mildly ($\sim$ 0.7 \AA) stronger with Mg-enhancement.
Furthermore, top panels of Figures 15 and 16 illustrate that the
broader $A$ indices (H$\gamma_{A}$, H$\delta_{A}$) show sensitivity to
Fe-enhancement and become significantly weaker with increasing age, by
up to 3 \AA\ at 12 Gyr for the case of the H$\delta_{A}$.  This is
mostly due to spectral effects, as can be seen from Table 5.  They are
illustrated in Figures 17 (H$\gamma$) to 18 (H$\delta$).  From this
experiment, it seems that for the high order Balmer lines, the $F$
indices (H$\gamma_{F}$, H$\delta_{F}$) are less prone to abundance
changes because of their narrower index definition and therefore may
possibly serve as more robust age indicators.  Furthermore, it is
found that H$\delta_{F}$ among Balmer lines has the least sensitivity
to C, N, and O (see also \citealt{sch02}, \citealt{pro07}).  

In Figure 19, we show some selected diagrams for illustrating element 
enhancement.  In the upper left panel, CN$_2$ and G4300 are compared 
and the effects of carbon and nitrogen enhancement are depicted.  The 
plot shows that CN$_2$ is sensitive to both carbon and nitrogen.  
CN$_2$ index values go up by 0.04 mag at 12 Gyr 
with 0.3 dex nitrogen enhancement here.  G4300, on the contrary, is 
primarily a carbon sensitive index.  Ca4227 and G4300 are 
contrasted in the lower left panel of Figure 19 and the effects of 
sulfur, calcium, and titanium are illustrated.  It is seen that Ca4227 
is significantly affected by calcium enhancement whereas G4300 is not
much altered by anything but carbon.  In the upper right panel, Mg $b$ 
and Fe5406 are compared and the effects of magnesium, and silicon are 
shown.  It is clear that Mg $b$ is notably affected by the magnesium 
enhancement.  H$\beta$ and Fe5406 are contrasted in the lower right 
panel and, as we have seen in Figures 10 and 13, H$\beta$ is not Fe-enhancement 
sensitive, while Fe5406 is significantly affected by Fe.
 
Figure 20 is basically same as Figure 19, but here the isochrone 
effects are decoupled from those of synthetic spectra.  The 
lines show the combination of isochrones and stellar spectral 
effects, while the points depict the isochrone effects alone 
at 1, 2, 5, and 12 Gyr.  It is seen again from Figure \ref{fig20} that 
the synthetic spectra make the spectral lines stronger and/or
weaker, while isochrones play a comparatively minor role.  The 
displacement of points in the bottom right panel is 
mostly due to the different [Fe/H] values that go into the index value 
calculations rather than the isochrone changes (see Figure 1 of Paper 
I).

\section{Summary and Discussion}  

We have commenced this project to make stellar population models which 
incorporate flexible chemistry, so that almost any interesting
chemical mixture can be interpolated.  Paper I dealt with the effects 
on the stellar evolution models, examining the temperature and 
luminosity changes due to the altered opacites when ten chemical 
elements are individually tweaked to the end of the red giant branch.  
In this paper we combine those isochrone effects with the stellar 
spectral effects in order to investigate 
their mixed effects on the integrated spectrophotometric indices as 
well as in their integrated spectra themselves.  We again emphasize 
here that the models in this study are {\em incomplete in terms of
inclusion of all stellar evolutionary phases and should not be used 
blindly when comparing to real stellar populations} until the 
helium-burning phases are properly incorporated.  A version with a 
full range of metallicity and with horizontal-branch and asymptotic 
giant branch stars included is planned.  Comparison of our models with 
observations of Virgo cluster galaxies will be presented as well 
(Serven et al. in preparation).

Within our spectral coverage (3000 \AA\ to 10000 \AA), we have 
investigated the broadband color behaviors in the $UBVR_CI_C$ filter 
set.  As one would expect, older populations show larger spectral 
effects due to element-by-element abundance changes.  This mostly 
reflects temperature effects.  We have also confirmed that Mg is 
the most important $\alpha$-element that shapes the $U-B$ and $B-V$ 
colors as already depicted in \citet{cas04}.

From our investigation of Lick indices using the \emph{integrated} spectra, 
we find that (1) CN$_2$ is a useful nitrogen indicator once we have 
good carbon abundances from G4300 and C$_2$4668, but good silicon (and 
also titanium for the C$_2$4668) abundances are also needed, (2) Ca4227 
is a robust calcium 
indicator with some good constraints of C, N, and O, (3) Fe4531 and Fe5015 
are very useful titanium indicators where an independent iron abundances 
is provided, (4) Mg $b$ and Fe5406 are good magnesium and iron indicator, 
respectively, and (5) the variation of individual elements affects the Balmer 
lines.  We defer the investigation of NaD and TiO$_1$ and 
TiO$_2$ indices until we have the 
Na-enhanced isochrones and can model TiO molecular effects 
with confidence.  Below, we illuminate some points 
that we have described here with the help of full spectra. 

Fe4531 and Fe5015: From Figure 9 we see that both Fe4531 and 
Fe5015 prove to be good titanium indicators.  They are almost 
equally sensitive to titanium and iron.  
Figures 21--23 show the SAURON \citep{bac01} 
spectral range, 4810 \AA\ to 5350 \AA, that includes H$\beta$, 
Fe5015, Mg$_1$, Mg $b$, Fe5270 and a part of Fe5335.  Here the 
integrated spectra of solar-scaled, Mg-, Fe-, and Ti-enhanced cases 
are shown at 2 Gyr with 300 km/sec velocity dispersion normalized at 
4750 \AA.  H$\beta$ is not influenced much by the enhancement 
of these elements at this age, as Figure 10 also shows.  
Mg $b$ is mostly sensitive to Mg, as Figure 21 shows, while 
Fe5015, Fe5270, and Fe5335 are notably sensitive to Fe-enhancement, 
although Fe5015 is equally sensitive to titanium, as Figures 22 and 23 display.  

Moreover, we have found that the Fe-enhanced and Ti-enhanced spectra 
show their centroids at different wavelengths in the case of Fe4531, 
opposite sides from the scaled-solar spectra, because of different 
locations of iron and titanium lines in their index bandpass.  Figure 
\ref{fig24} displays this case\footnote{ This phenomenon 
at the stellar level with higher resolution can be seen at  
http://astro.wsu.edu/hclee/Fe45$\_$4000mp00gp25$\_$ti.pdf  and 
http://astro.wsu.edu/hclee/Fe45$\_$4000mp00gp25$\_$fe.pdf}.  
We have preliminary hints from high-S/N, high-resolution spectra of 
Virgo cluster galaxies taken at the Kitt Peak 
4-m telescope (Serven et al. in preparation) 
that galaxies with the higher velocity dispersion tend to have their 
centroids at the longer wavelengths, indicating higher 
Ti-enhancements.  If confirmed, this would indicate a Ti-$\sigma$ relation, 
similar to the well-known Mg-$\sigma$ relation (see also \citealt{mil00}).

Balmer lines:

H$\beta$: The bottom right panel of Figure 10 shows that, in an 
$\alpha$-enhanced mixture at the fixed [Fe/H] (Alpha+), H$\beta$ becomes 
weaker.  This sounds a bit different from the conclusion of LW05 where 
H$\beta$ is $\alpha$-element insensitive.  As we can see from Table 5, the 
effects that we see from the bottom right panel of Figure 10 are mostly 
isochrone effects, especially at young ages.  The H$\beta$ index
becomes weaker because of the decrease of temperature of stellar 
isochrones from the $\alpha$-enhanced case at fixed [Fe/H] (see also 
\citealt{coe07}; \citealt{sch07}).  LW05 used only solar-scaled isochrones 
at solar and super-solar metallicities, although these authors employed 
modified stellar spectra because of the $\alpha$-enhancement.  

Moreover, it is worth emphasizing that $\alpha$-enhanced isochrones 
at fixed $Z$ are now no longer significantly hotter than the 
solar-scaled ones.  This has been revealed by Paper I, in its 
figure 11, after incorporating the latest development of low-temperature 
opacities by \citet{fer05}.  If this behavior is indeed true, it will certainly 
affect the conclusions of earlier works (e.g., \citealt{tm03})
that are based upon the \citet{sal00} isochrones, which were later 
shown by \citet{wei06} to present problems.

H$\gamma$ and H$\delta$: Fe-enhancement makes these indices 
significantly weaker, especially for the broader indices, H$\gamma_A$ 
and H$\delta_A$ (see also \citealt{gs08}).  Figures 17 and 18 show that 
this is because of Fe-lines that inhabit the continuum region.  It is 
suggested therefore that the narrower indices (H$\gamma_F$ and 
H$\delta_F$) may serve as better age indicators in case the stellar 
systems are of wildly different chemical mixtures.

Our goals are to narrow down the uncertainties of mean age estimations 
to 10\% when derived from a single integrated light spectrum.  We 
believe that we can reach these goals as we are able to take into 
account the effects from individual chemical elements.  The ultimate 
goal then obviously would be the understanding of the chemical 
enrichment and star formation histories of galaxies, stepping away 
from the study of mean ages and mean metallicities to begin to tackle 
the more realistic problem of multi-age, multi-metallicity stellar 
populations as they exist in galaxies.  We believe that several efforts such as 
SPECKMAP \citep{ocv06}, MOPED \citep{pan07}, and STARLIGHT 
\citep{cid08} will be even more useful once they use sophisticated 
stellar population models with flexible chemistry as inputs.

\acknowledgements 
We thank Jim Rose, Jon Fulbright, and Andreas Koch for useful discussions 
on the Mg-enhancement of the metal-rich cool giants.  We are also grateful 
to the referee, Ricardo Schiavon, for a thorough reading which greatly 
helped us to improve the paper.  
Support for this work was provided by the NSF through grant AST-0307487, 
the New Standard Stellar Population Models (NSSPM) project.  H.-c. Lee also 
acknowledges JP Blakeslee's support via NASA grant.  PC acknowledges the support 
by the European Community under a Marie Curie International Incoming Fellowship.


\begin{deluxetable}{lrrrrrrr}
\tablecolumns{8}
\tabletypesize{\scriptsize}
\tablecaption{Gross Color Behavior With Metallicity Increase \label{tab:color1}}
\tablehead{
\colhead{Color} & \colhead{Synthetic } & \colhead{$\Delta$
Color,} & \colhead{$\Delta$ Color,} & \colhead{$\Delta$ Color} & \colhead{Empirical} & \colhead{$\Delta$
Color, } & \colhead{$\Delta$ Color,  } \\
   & \colhead{Color} & \colhead{24 Elements- } & \colhead{All Elements- }  & \colhead{$T_{\rm eff}$} & \colhead{Color} & \colhead{All Elements- } & \colhead{$T_{\rm eff}$ } \\
   &  & \colhead{enhanced } & \colhead{enhanced }  &  &  & enhanced  &  \\
 & \colhead{(mag)} & \colhead{(mmag)} & \colhead{(mmag)} & \colhead{(mmag)} & \colhead{(mag)} & \colhead{(mmag)} & \colhead{(mmag)}  }

\startdata

\sidehead{Dwarf, $T_{\rm eff}=4000$ K} 

$U-B$ & 1.26 &  21 &  81 & -106 & $ 1.36\pm 0.07 $ &  58 &  -85 \\
$B-V$ & 1.43 &  43 &  35 & -137 & $ 1.36\pm 0.02 $ &  22 &  -81 \\
$V-R$ & 0.89 &  16 &  24 &  -94 & $ 0.82\pm 0.01 $ &   3 &  -74 \\
$V-I$ & 1.60 &  13 &  24 & -176 & $ 1.58\pm 0.02 $ &  -5 & -176 \\

\sidehead{Dwarf, $T_{\rm eff}=5000$ K}

$U-B$ & 0.61 & 123 & 111 & -179 & $ 0.71\pm 0.07 $ & 160 & -169 \\
$B-V$ & 0.91 &  55 &  45 &  -90 & $ 0.96\pm 0.02 $ &  62 & -101 \\
$V-R$ & 0.52 &  21 &  19 &  -62 & $ 0.52\pm 0.01 $ &  11 &  -55 \\
$V-I$ & 0.98 &  15 &  14 & -103 & $ 0.96\pm 0.02 $ &  13 &  -97 \\
\enddata
\tablecomments{1. The ``Dwarf'' is $\log g = 4.5$, and the ``Giant'' is $\log g = 2.5$.  
               2. Columns 2 presents the calculations at [R/H] = 0, where R is 
               the generic heavy element (see section 2.2).  3. Column 3 presents the 
               case where 24 elements (C, N, O, Na, Mg, Al, Si, S, Cl, K, Ca, Sc, 
               Ti, V, Cr, Mn, Fe, Co, Ni, Cu, Zn, Sr, Ba, Eu) are enhanced by 0.3 dex.  
               4. Columns 4 and 7 present the calculations at [R/H] = 0.3 dex.  
               5. Columns 5 and 8 present the color changes when the temperature is 
               increased by 250 K, synthetic and empirical, respectively.}
\end{deluxetable}

\begin{deluxetable}{lrrrrrrrrrrr}
\tablecolumns{12}
\tabletypesize{\scriptsize}
\tablecaption{Color Behavior With Enhancements At Constant $Z$ \label{tab:color2}}
\tablehead{
\colhead{Color} & \colhead{C} 
& \colhead{N}  & \colhead{O} 
& \colhead{Ne} & \colhead{Mg} & \colhead{Si}
& \colhead{S}  & \colhead{Ca} & \colhead{Ti} & \colhead{Fe} 
& \colhead{$\alpha$} \\  }

\startdata

\sidehead{Dwarf, $T_{\rm eff}=4000$ K}

$U-B$ &   -13 &    8 &   19 &   -2 &  -66 &  -36 &    0 & -40 &   15 &  136 & -103 \\
$B-V$ &    17 &   -2 &  -50 &   -4 &  -19 &   35 &    0 &  -7 &   15 &   22 &  -33 \\
$V-R$ &    -2 &   -1 &   -7 &   -1 &   31 &   -1 &    0 & -14 &    4 &   12 &    8 \\
$V-I$ &    -5 &   -1 &   -5 &   -1 &   31 &   -1 &    0 & -14 &    4 &   14 &   10 \\

\sidehead{Dwarf, $T_{\rm eff}=5000$ K}

$U-B$ &    13 &   18 &  -61 &  -13 &  -26 &  -39 &   -2 &  -4 &   11 &   61 & -120 \\
$B-V$ &    17 &    1 &  -40 &   -6 &  -32 &    5 &   -1 &   6 &    7 &   18 &  -58 \\
$V-R$ &    -3 &   -5 &   -8 &   -2 &    8 &   -2 &    0 &   0 &    3 &   10 &   -3 \\
$V-I$ &   -12 &  -11 &    0 &   -1 &   10 &   -1 &    0 &  -1 &    3 &   15 &    7 \\
\enddata
\tablecomments{1. All elements scaled individually by +0.3 dex, 
                  except C, which is increased by +0.15 dex.  
               2. Numbers are in milli-magnitude.}
\end{deluxetable}

\begin{deluxetable}{lrrrrrrrrrrr}
\tablecolumns{12}
\tabletypesize{\scriptsize}
\tablecaption{Color Behavior With Enhancements At Constant [R/H] \label{tab:color3}}
\tablehead{
\colhead{Color} & \colhead{C} 
& \colhead{N}  & \colhead{O} 
& \colhead{Ne} & \colhead{Mg} & \colhead{Si}
& \colhead{S}  & \colhead{Ca} & \colhead{Ti} & \colhead{Fe} 
& \colhead{$\alpha$} \\  }

\startdata

\sidehead{Dwarf, $T_{\rm eff}=4000$ K}

$U-B$ &   -11 &   10 &   30 & \nodata &  -65 &  -35 &    0 &  -40 &   15 &  139 &  -97 \\
$B-V$ &    22 &    0 &  -31 & \nodata &  -17 &   38 &    0 &   -7 &   15 &   26 &   -5 \\
$V-R$ &     0 &    0 &    0 & \nodata &   31 &    0 &    0 &  -13 &    4 &   14 &   19 \\
$V-I$ &    -3 &    0 &    1 & \nodata &   32 &   -1 &    0 &  -14 &    4 &   16 &   19 \\

\sidehead{Dwarf, $T_{\rm eff}=5000$ K}

$U-B$ &    28 &   26 &  -10 & \nodata &  -20 &  -33 &    0 &   -3 &   11 &   74 &  -57 \\
$B-V$ &    24 &    4 &  -14 & \nodata &  -30 &    8 &    0 &    6 &    7 &   24 &  -23 \\
$V-R$ &    -1 &   -3 &    2 & \nodata &    9 &   -1 &    0 &    0 &    3 &   12 &   11 \\
$V-I$ &   -10 &  -10 &    7 & \nodata &   11 &    0 &    0 &   -1 &    3 &   16 &   18 \\
\enddata
\tablecomments{1. All elements scaled individually by +0.3 dex, 
                  except C, which is increased by +0.15 dex.  
               2. Numbers are in milli-magnitude.}
\end{deluxetable}

\begin{deluxetable}{lrrrrrrrrrrrr}
\tablecolumns{13}
\tabletypesize{\scriptsize}
\setlength{\tabcolsep}{0.05in}
\tablecaption{Model Population Color Behavior With Enhancements At Constant $Z$ \label{tab:color4}}
\tablehead{
\colhead{} & \colhead{Synthetic} &  \colhead{$\Delta$I} 
& \colhead{$\Delta$I}  & \colhead{$\Delta$I} 
& \colhead{$\Delta$I} & \colhead{$\Delta$I} & \colhead{$\Delta$I}
& \colhead{$\Delta$I}  & \colhead{$\Delta$I} & \colhead{$\Delta$I} & \colhead{$\Delta$I} 
& \colhead{$\Delta$I} \\ 
\colhead{Color} & \colhead{Color} &  \colhead{(C)} 
& \colhead{(N)}  & \colhead{(O)} 
& \colhead{(Ne)} & \colhead{(Mg)} & \colhead{(Si)}
& \colhead{(S)}  & \colhead{(Ca)} & \colhead{(Ti)} & \colhead{(Fe)} 
& \colhead{($\alpha$)} \\ 
\colhead{} & \colhead{(mag)} &  \colhead{(mmag)} 
& \colhead{(mmag)}  & \colhead{(mmag)} 
& \colhead{(mmag)} & \colhead{(mmag)} & \colhead{(mmag)}
& \colhead{(mmag)}  & \colhead{(mmag)} & \colhead{(mmag)} & \colhead{(mmag)} 
& \colhead{(mmag)} \\  }

\startdata

\sidehead{Population Age 1 Gyr}
$BC_V$ &  \nodata &  -0.3 &   0.0 &    5.5 &  1.9 &   2.9 &   2.6 &   1.6 &
-1.2 &   0.8 &  -3.8 & 8.0 \\
$U-B $ &    0.050 &  -3.7 &   2.3 &  -14.9 & -4.0 &   1.6 &  -5.5 &  -1.9 &
2.6 &   2.2 &  14.8 & -20.7 \\
$B-V $ &    0.361 &  -1.0 &  -0.2 &  -34.6 &  1.3 &   5.7 &   9.1 &   6.2 &
6.0 &   2.4 &  -1.1 & -11.6 \\
$V-R $ &    0.223 &  -1.1 &  -0.2 &  -17.1 &  2.4 &   7.6 &   7.4 &   4.8 &
1.0 &   1.9 &  -1.7 & -2.3 \\
$V-I $ &    0.458 &  -4.0 &   0.4 &  -26.3 &  5.8 &  15.2 &  15.2 &  10.7 &
2.0 &   7.2 &  -7.6 & 4.5 \\


\sidehead{ Population Age   2 Gyr }
$BC_V$  &  \nodata& -9.9 & -4.5 & 11.1 & 0.7 &  12.2 &   4.5 &  0.6 & -17.5 &  15.5 &   3.9 &   46.6 \\
$U-B$   &  0.171  &  0.5 & 12.1 &-30.2 & 4.9 &   7.7 &  -6.6 &  4.8 &  -2.8 &  16.7 &  12.5 &  -18.4 \\
$B-V$   &  0.664  & 16.1 &  6.3 &-20.6 &12.9 & -10.0 &  10.4 &  8.2 &  -2.4 &  15.2 &  -2.5 &  -10.4 \\
\enddata
\tablecomments{1. All elements scaled individually by +0.3 dex, 
                  except C, which is increased by +0.2 dex (see footnote \#10).}
\end{deluxetable}

\clearpage
\LongTables
\begin{landscape}
\begin{deluxetable}{lrrrrrrrrrrrrrrr}
\tablecolumns{16}
\tabletypesize{\scriptsize}
\setlength{\tabcolsep}{0.04in}
\tablecaption{Model Population Index Behavior With Element Enhancements  \label{tab:index5}}
\tablehead{
\colhead{Index} & \colhead{Index} &  \colhead{$\Delta$I} 
& \colhead{$\Delta$I}  & \colhead{$\Delta$I} 
& \colhead{$\Delta$I} & \colhead{$\Delta$I} & \colhead{$\Delta$I}
& \colhead{$\Delta$I}  & \colhead{$\Delta$I} & \colhead{$\Delta$I}
& \colhead{$\Delta$I} & \colhead{$\Delta$I} & \colhead{$\Delta$I} &
\colhead{$\Delta$I} 
& \colhead{$\Delta$I} \\ 
\colhead{Name} & \colhead{Value} &  \colhead{(C)} & \colhead{(C+)}  
& \colhead{(N)}  & \colhead{(N+)} & \colhead{(O)} & \colhead{(O+)} 
& \colhead{(Mg)} & \colhead{(Si)}
& \colhead{(S)}  & \colhead{(Ca)} & \colhead{(Ti)} & \colhead{(Fe)} 
& \colhead{($\alpha$)} & \colhead{($\alpha$+)} \\
}

\startdata

\sidehead{ Population Age 1 Gyr, Isochrone Effects Only}
      CN$_1$   &   -0.1729 & -0.0013 &  0.0052 & -0.0003 &  0.0015 & -0.0091 &  0.0096 &  0.0029 &  0.0027 &  0.0014 &  0.0004 & -0.0001 &  0.0037 & -0.0030 &  0.0317 \\
      CN$_2$   &   -0.1099 & -0.0008 &  0.0031 &  0.0001 &  0.0009 & -0.0052 &  0.0061 &  0.0020 &  0.0017 &  0.0010 &  0.0005 &  0.0000 &  0.0017 & -0.0014 &  0.0210 \\
      Ca4227   &    0.3702 & -0.0054 &  0.0184 & -0.0032 &  0.0029 & -0.0492 &  0.0247 &  0.0100 &  0.0098 &  0.0061 &  0.0018 & -0.0003 &  0.0391 & -0.0349 &  0.1102 \\
       G4300   &   -0.1319 & -0.0663 &  0.2024 & -0.0232 &  0.0580 & -0.4061 &  0.3611 &  0.1018 &  0.0976 &  0.0473 &  0.0105 & -0.0048 &  0.2476 & -0.2056 &  1.1606 \\
      Fe4383   &    0.0769 & -0.0418 &  0.0922 & -0.0086 &  0.0337 & -0.2781 &  0.1782 &  0.0566 &  0.0289 &  0.0066 &  0.0178 &  0.0081 &  0.2371 & -0.2220 &  0.7064 \\
      Ca4455   &    0.4773 & -0.0219 &  0.0285 & -0.0098 &  0.0062 & -0.1185 &  0.0428 &  0.0098 &  0.0069 &  0.0040 &  0.0023 & -0.0006 &  0.1553 & -0.1151 &  0.1653 \\
      Fe4531   &    1.4918 & -0.0271 &  0.0517 & -0.0082 &  0.0124 & -0.1626 &  0.0814 &  0.0261 &  0.0209 &  0.0124 &  0.0067 &  0.0005 &  0.1565 & -0.1401 &  0.3023 \\
  C$_2$ 4668   &    0.3137 & -0.0962 &  0.0786 & -0.0405 &  0.0117 & -0.4479 &  0.1322 &  0.0245 &  0.0041 &  0.0059 &  0.0118 & -0.0001 &  0.7050 & -0.4688 &  0.5675 \\
\enddata
\tablecomments{1. The units of CN$_1$, CN$_2$, Mg$_1$, Mg$_2$, TiO$_1$, TiO$_2$ are 
               mag.  That of the rest are \AA.  2. Index Value in the second column
               is the Index Value of solar-scaled case.  3. $\Delta$ I = Index Value of 
               each element-enhanced case $-$ Index Value of solar-scaled.  4. C+, N+, O+, 
               and $\alpha$+ are those element-enhanced cases at fixed [Fe/H], whereas
               the rest are element-enhanced cases at fixed $Z$.  5. C and C+ are 
               0.2 dex enhanced cases and $\alpha$+ is 0.4 dex enhanced case, while 
               the rest of cases are 0.3 dex enhanced cases.  6. $\alpha$-elements are 
               O, Ne, Mg, Si, S, Ca, and Ti in this study.}
\end{deluxetable}
\clearpage
\end{landscape}


\clearpage

\begin{figure}
\epsscale{1.}
\plotone{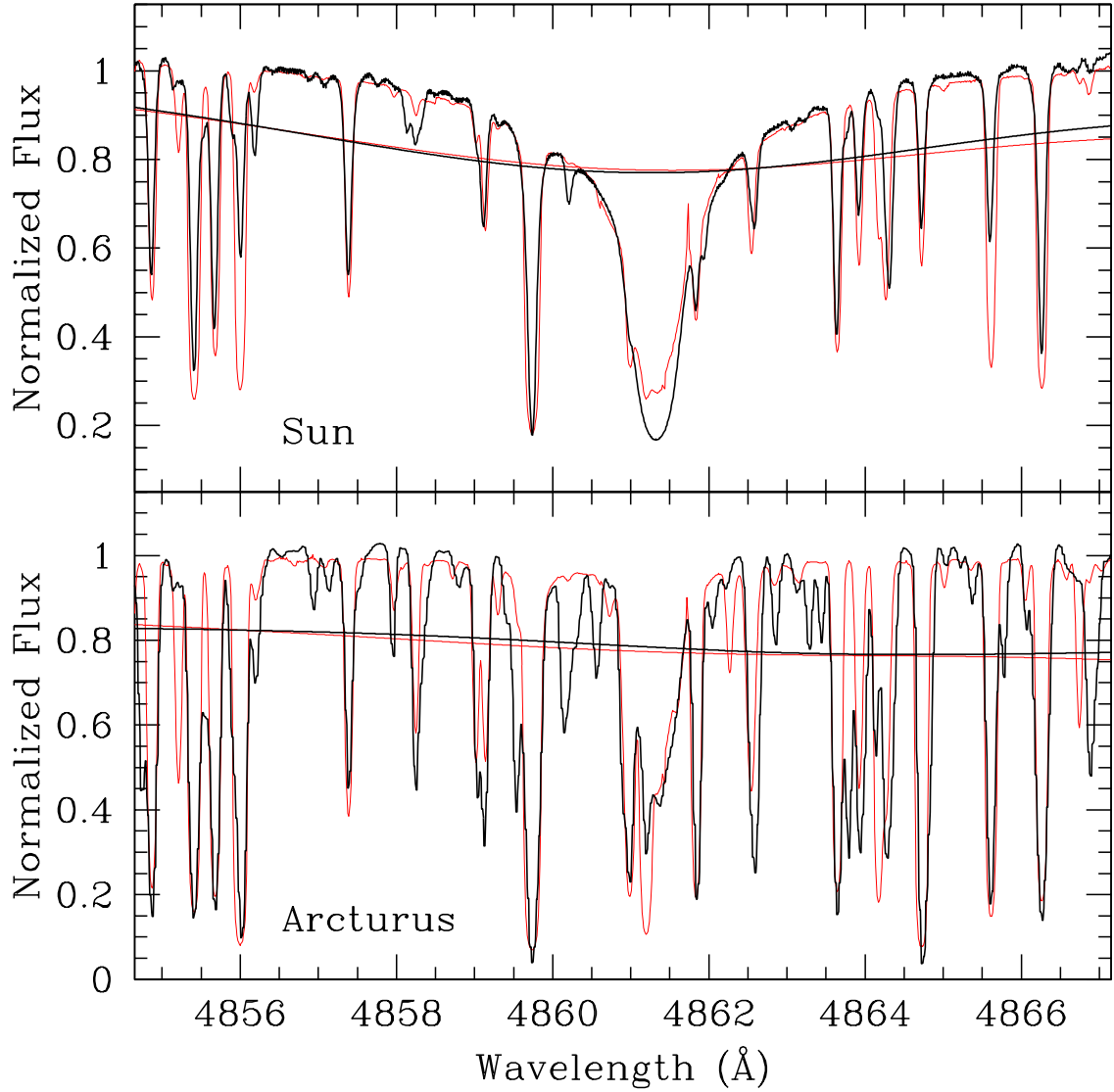}
\caption{The synthetic and observed spectra are compared near the 
region of the H$\beta$ lines.  The top panel is the solar spectra and 
the bottom panel is Arcturus one. The thicker (black) lines are the observations.  
The degraded low-resolution spectra at 200 km/sec are overlaid on top 
of the high-resolution spectra.  It is seen that the notable mismatches 
between the synthetic and the observed spectra at the high-resolution 
comparisons are smeared out at the low-resolution.  \label{fig01}}
\end{figure}

\begin{figure}
\epsscale{1.}
\plotone{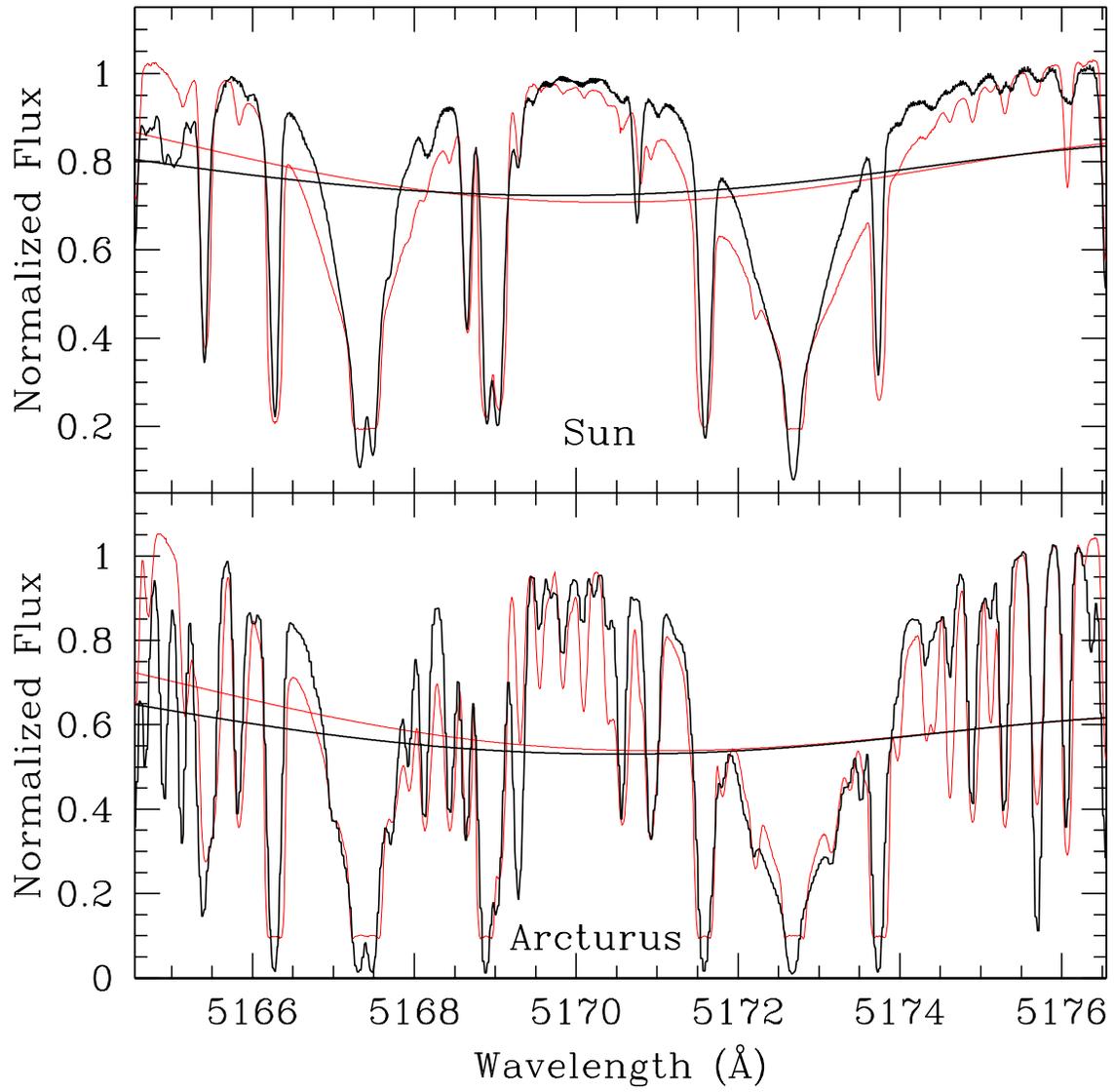}
\caption{Similar to Figure 1, but near the region of the 
Mg $b$ lines.  \label{fig02}}
\end{figure}

\begin{figure}
\epsscale{1.}
\plotone{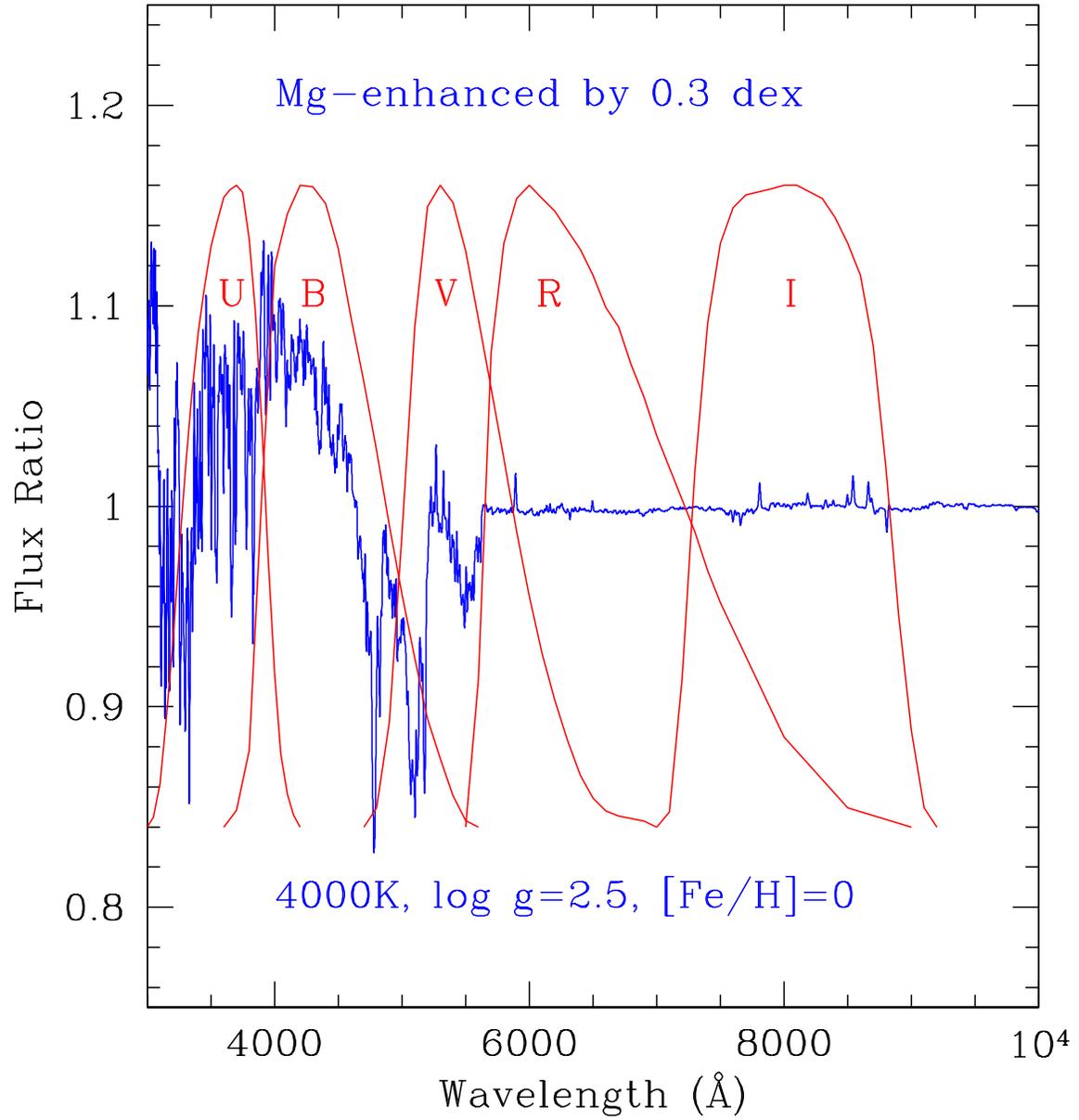}
\caption{At 4000 K and log g = 2.5, solar-scaled and Mg-enhanced spectra 
are divided.  $UBVRI$-band filter response curves are overlaid.  The blue colors 
of $U - B$ and $B - V$ and red colors of $V - R$ and  $V - I$ from the 
Mg-enhanced spectra that we read in Table 3 are visually elucidated here.  
\label{fig03}}
\end{figure}

\begin{figure}
\epsscale{1.}
\plotone{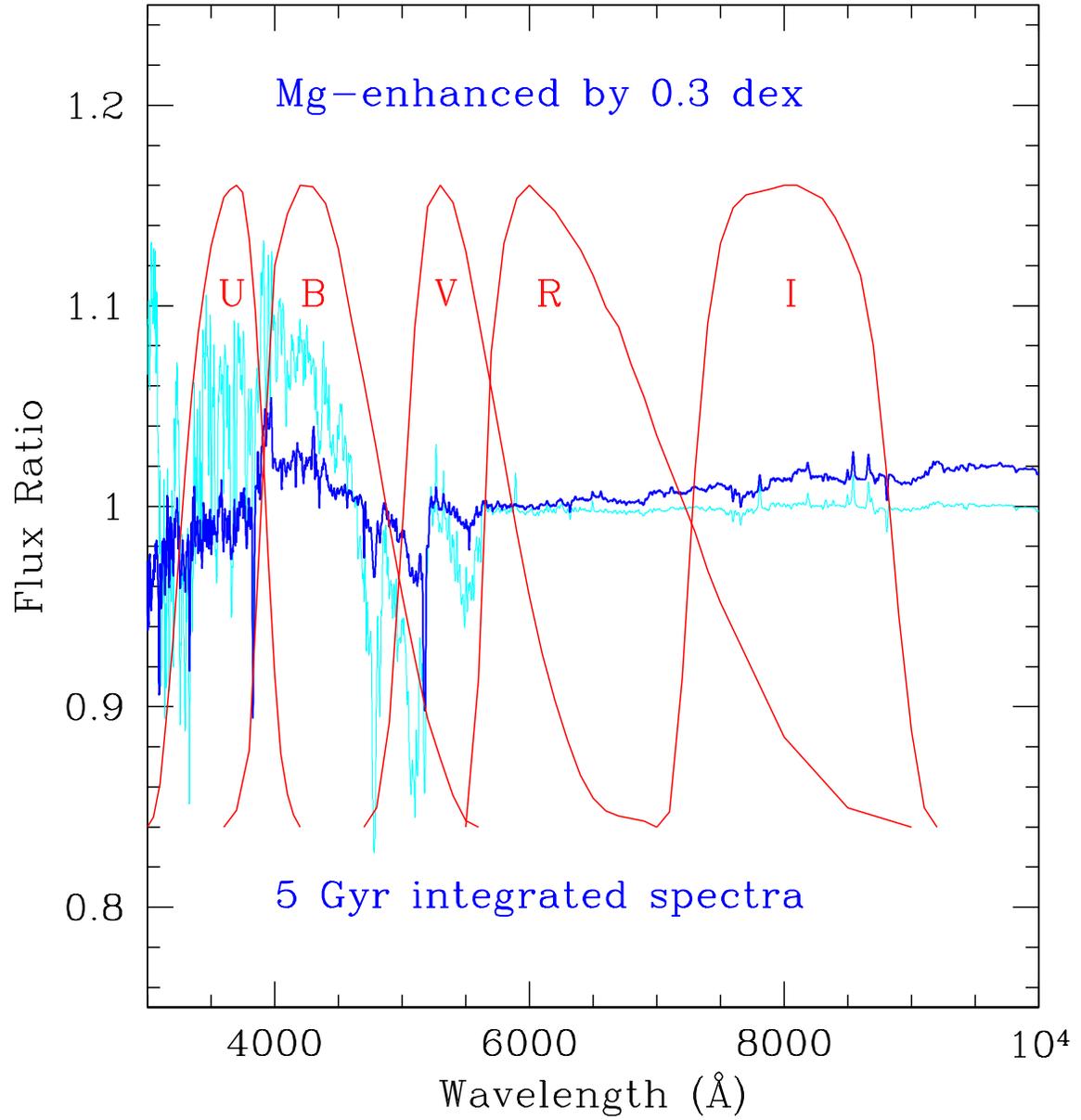}
\caption{Similar to Figure 3, but here the thicker line is the flux ratio 
between the solar-scaled and the Mg-enhanced spectra of the 5 Gyr integrated 
spectra at solar metallicity.  It is normalized at 6000 \AA.  The color 
changes because of the Mg-enhancement that we read from Table 4 can be 
vividly seen here (see text).  \label{fig04}}
\end{figure}

\begin{figure}
\epsscale{1.}
\plotone{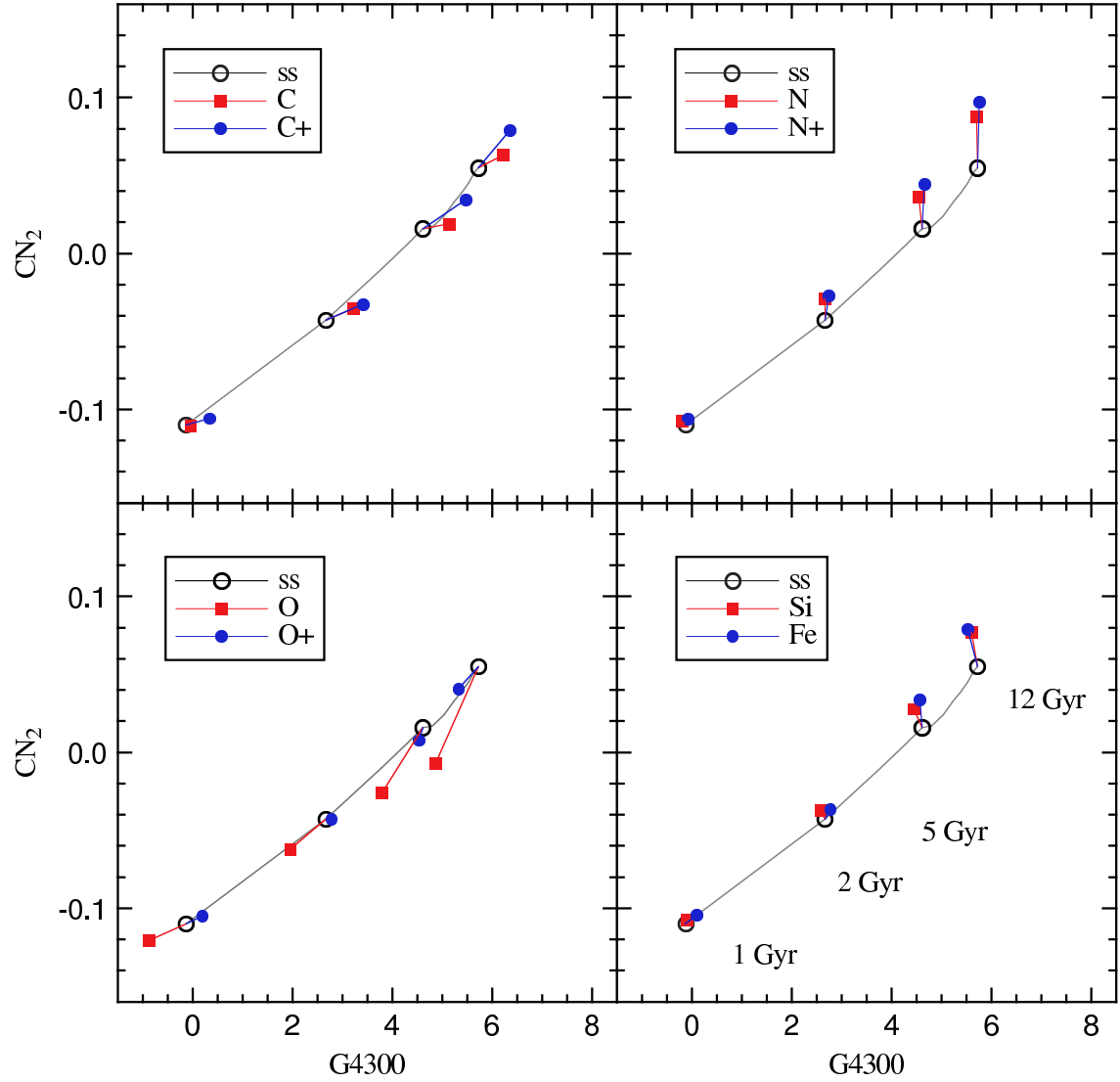}
\caption{CN$_2$ is plotted as a function of G4300 from 1 Gyr 
to 12 Gyr for simple stellar populations at solar metallicity.  
Carbon-enhanced (upper left), nitrogen-enhanced (upper right), 
oxygen-enhanced (lower left), and silicon-, iron-enhanced cases (lower 
right) at fixed $Z$ are depicted at 1, 2, 5, and 12 Gyr.  C+, N+, 
and O+ indicate the carbon-, nitrogen-, and oxygen-enhanced 
cases at fixed [Fe/H], respectively. \label{fig05}}
\end{figure}

\begin{figure}
\epsscale{1.}
\plotone{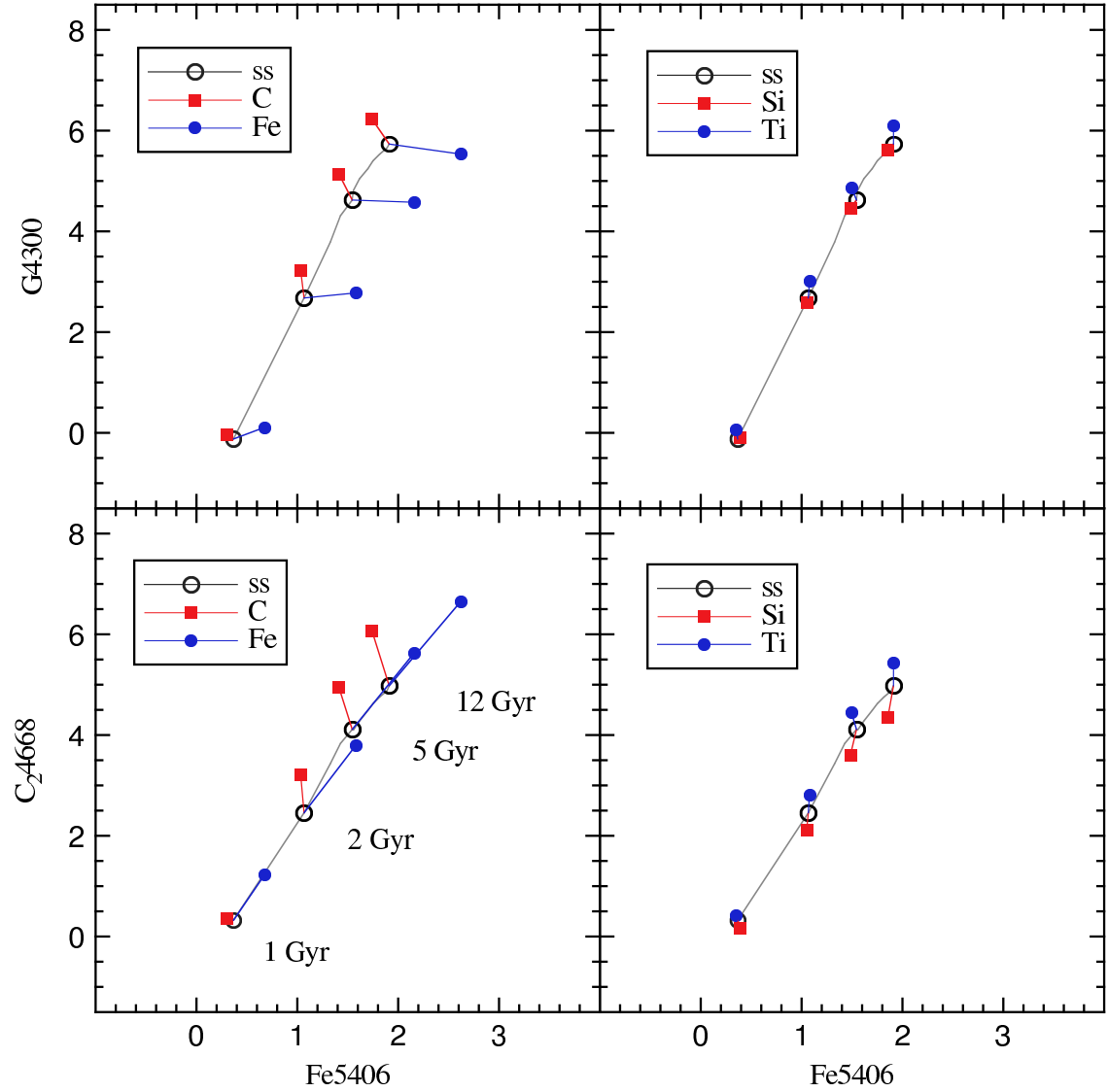}
\caption{Fe5406 vs. G4300 (top) and C$_2$4668 (bottom).  Left panels 
show carbon- and iron-enhanced cases, while right panels display 
silicon- and titanium-enhanced cases.  Note that C$_2$4668 is 
significantly affected both by carbon and iron.  C$_2$4668 is further 
sensitive to silicon- and titanium-enhancement. \label{fig06}}
\end{figure}

\begin{figure}
\epsscale{1.}
\plotone{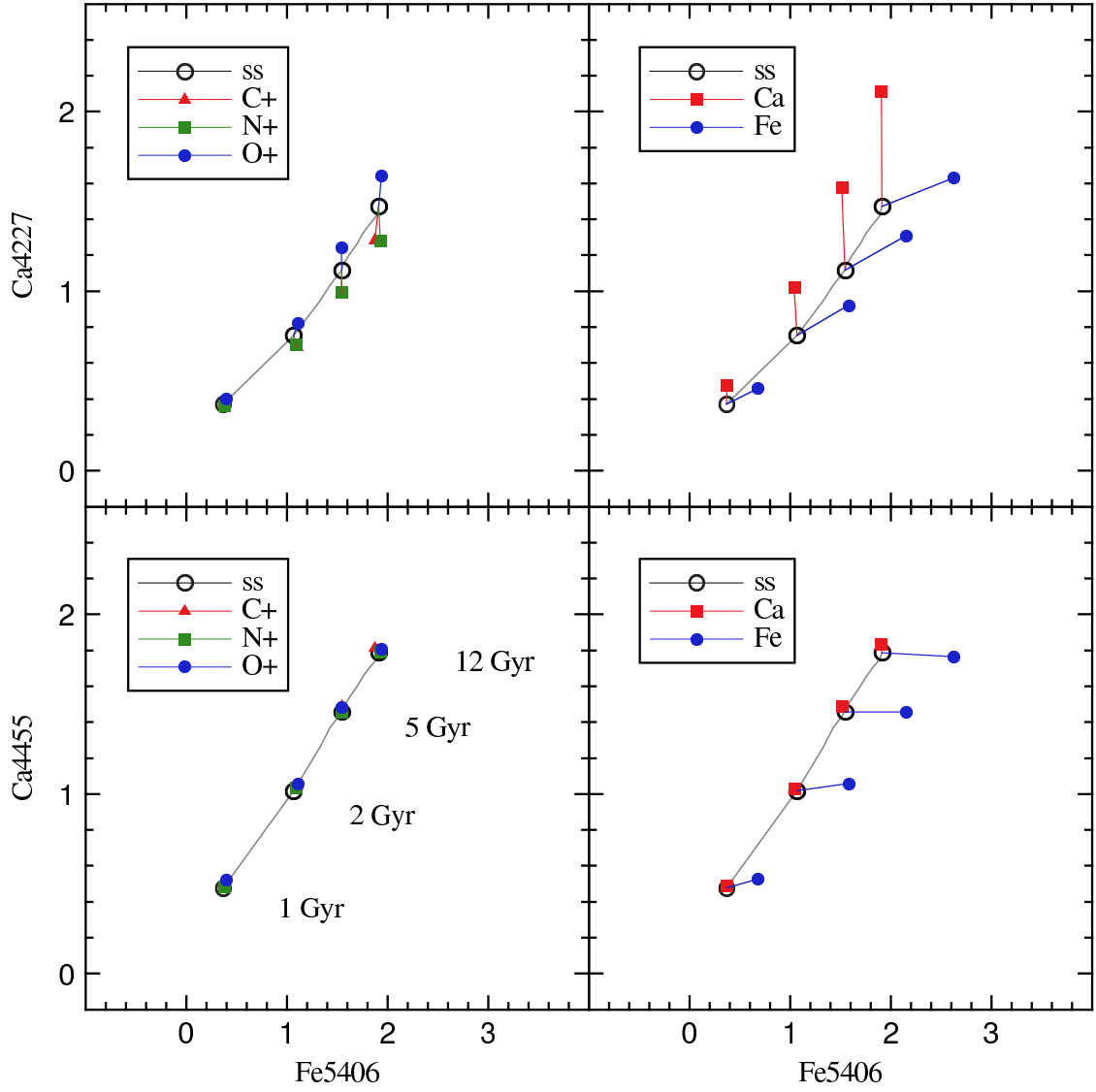}
\caption{Fe5406 vs. Ca4227 (top) and Ca4455 (bottom).  Left panels 
show carbon-, nitrogen-, and oxygen-enhanced cases at fixed [Fe/H], 
while right panels depict calcium-, iron-enhanced cases at fixed $Z$.  
Ca4227 is predominantly calcium-sensitive as well as 
a contribution from C, N, O, and Fe, while Ca4455 is hardly 
altered by those elements.\label{fig07}}
\end{figure}

\begin{figure}
\epsscale{1.}
\plotone{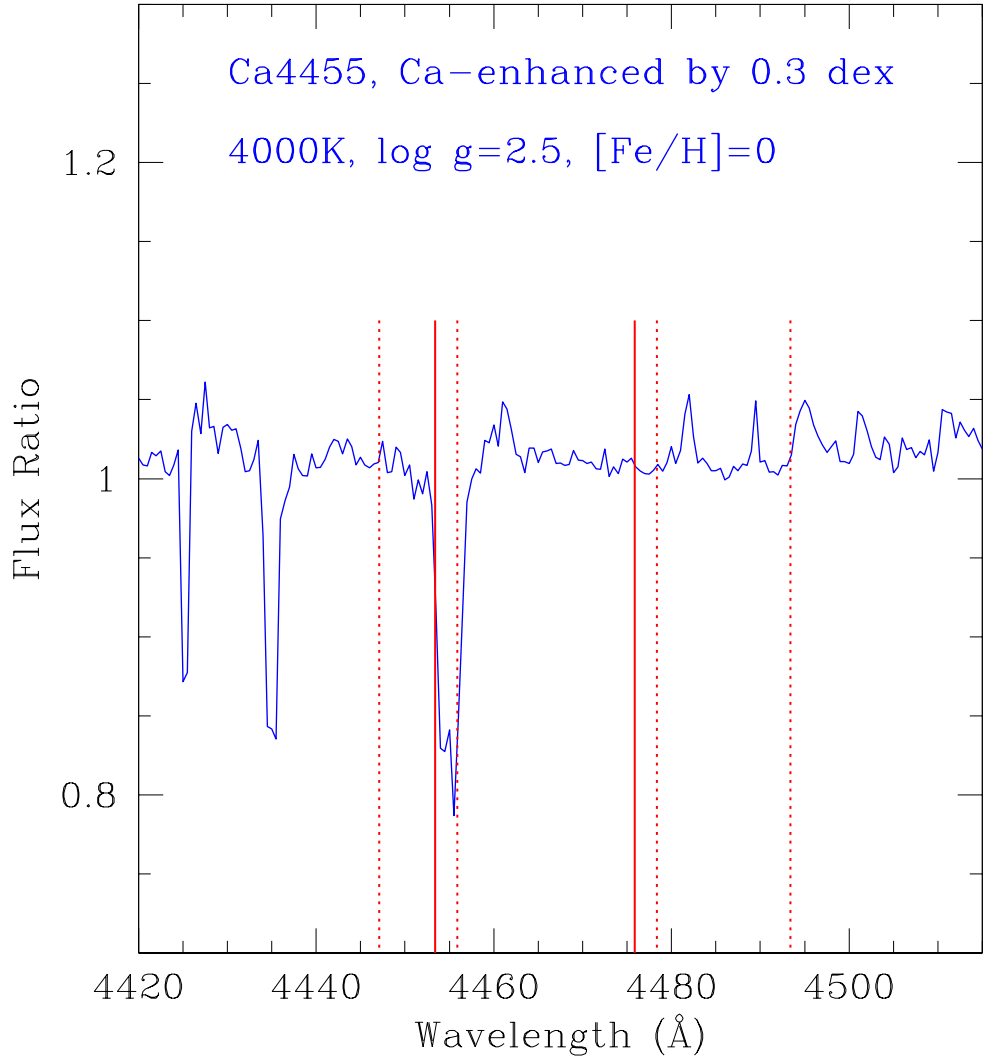}
\caption{At 4000 K and log g = 2.5, solar-scaled and Ca-enhanced spectra 
are divided.  The index definition of \citet{wor94b} for the Ca4455 index is 
depicted with straight lines (solid lines for the index bandpass and dotted 
lines for the pseudo-continua, respectively).  Note that the strong 
Ca line feature is shared both by the index bandpass and the blue continuum 
that cancel out the calcium effect on the Ca4455 index. \label{fig08}}
\end{figure}

\begin{figure}
\epsscale{.6}
\plotone{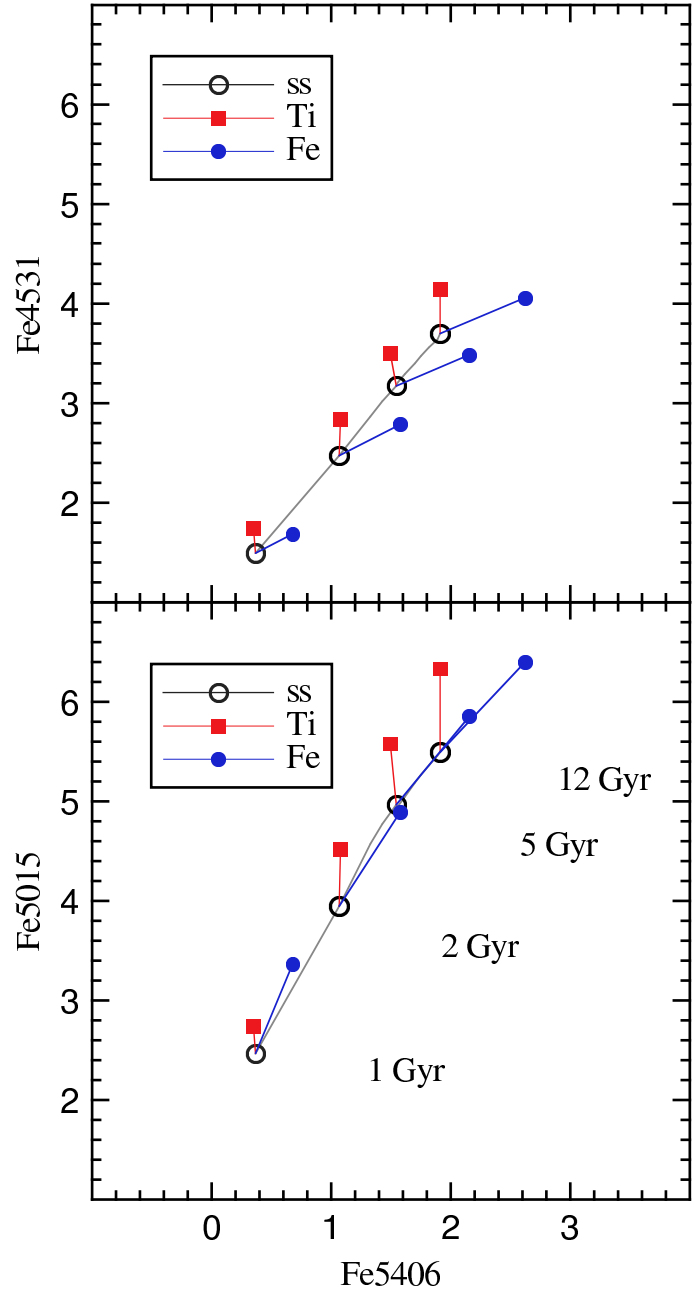}
\caption{Fe5406 vs. Fe4531 (top) and Fe5015 (bottom).  Titanium- and 
iron-enhanced cases are depicted.  Both of these Lick indices show 
a titanium sensitivity equal to that of iron. \label{fig09}}
\end{figure}

\begin{figure}
\epsscale{1.}
\plotone{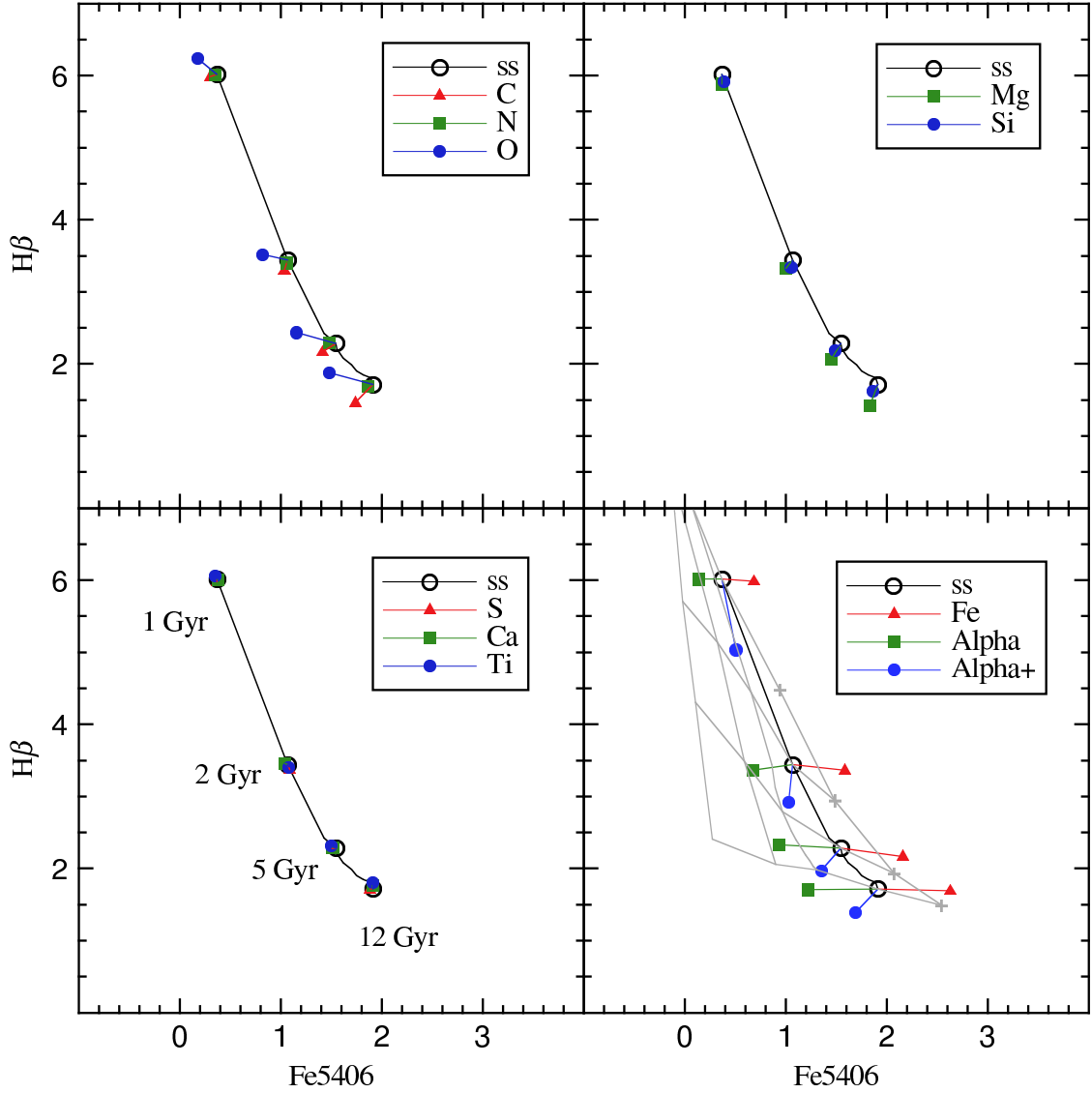}
\caption{Displacements are shown in diagrams of H$\beta$ as a 
function of Fe5406 as each chemical element is enhanced by 0.3 dex, 
except carbon, which is enhanced by 0.2 dex.  In the lower right 
panel, solar-scaled chemical mixtures of [Fe/H] = $-$2.0, 
$-$1.0, $-$0.5, 0.0, and 0.5 are connected at 1, 2, 5, and 12 Gyr.  To
guide the eye, the [Fe/H] = 0.5 line is marked with plus
signs at given ages.  Also in the lower right panel, the
$\alpha$-elements enhanced case at fixed [Fe/H] is shown (see
text). \label{fig10}}
\end{figure}

\begin{figure}
\epsscale{.7}
\plotone{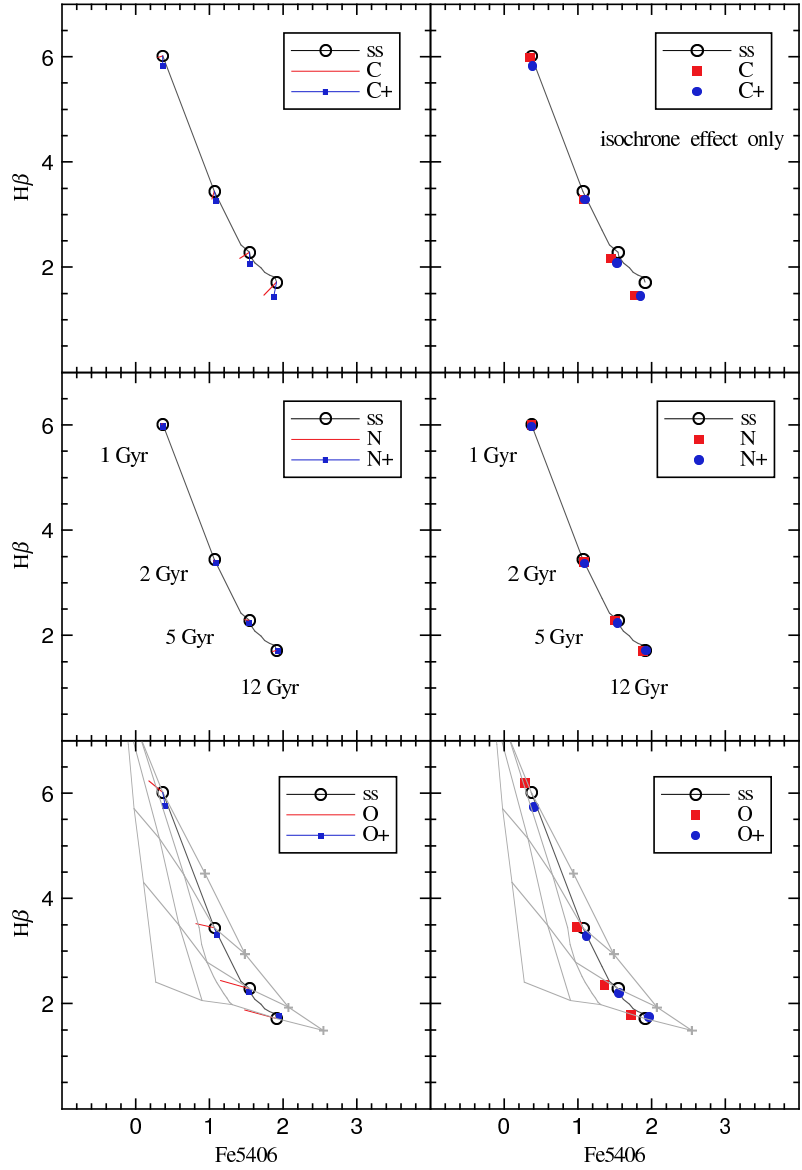}
\caption{C-, N-, O-enhanced cases are shown from top to bottom 
in diagrams of H$\beta$ as a function of Fe5406.  The same 
solar-scaled chemical mixture grids with a range of metallicity and 
age are displayed here in the bottom panels as in the
bottom panels of Figure 10.  The left panels show the combination of 
isochrone and stellar spectral effects, while the right panels 
display the isochrone effects alone. \label{fig11}}
\end{figure}

\begin{figure}
\epsscale{1.}
\plotone{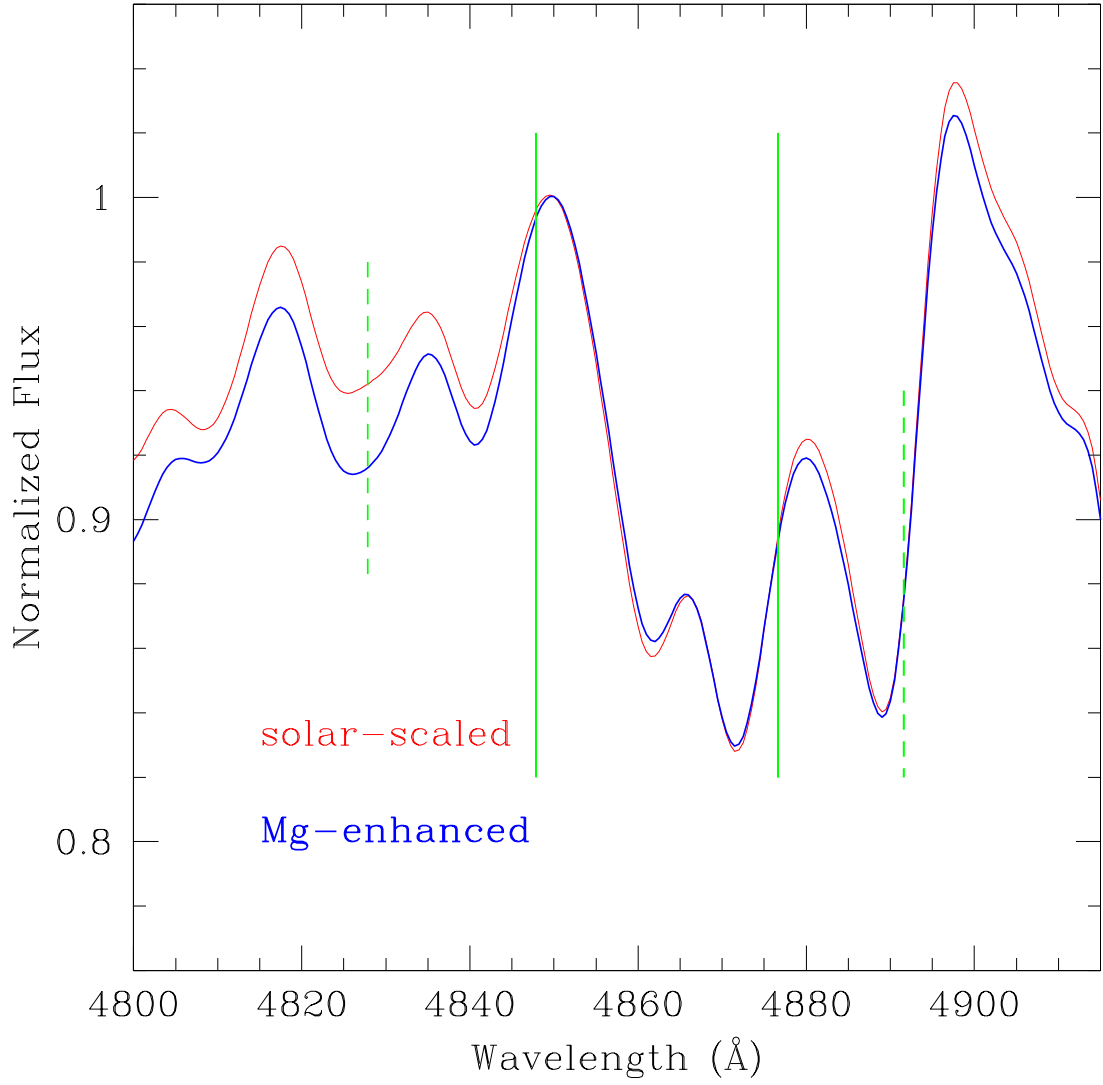}
\caption{Solar-scaled and Mg-enhanced 12 Gyr integrated spectra near 
H$\beta$ (4861 \AA).  Note that the Mg-lines make the blue-continuum levels 
significantly lower, which consequently make the H$\beta$ index strengths 
weaker as we see in the upper right panel of Figure 10. \label{fig12}}
\end{figure}

\begin{figure}
\epsscale{1.}
\plotone{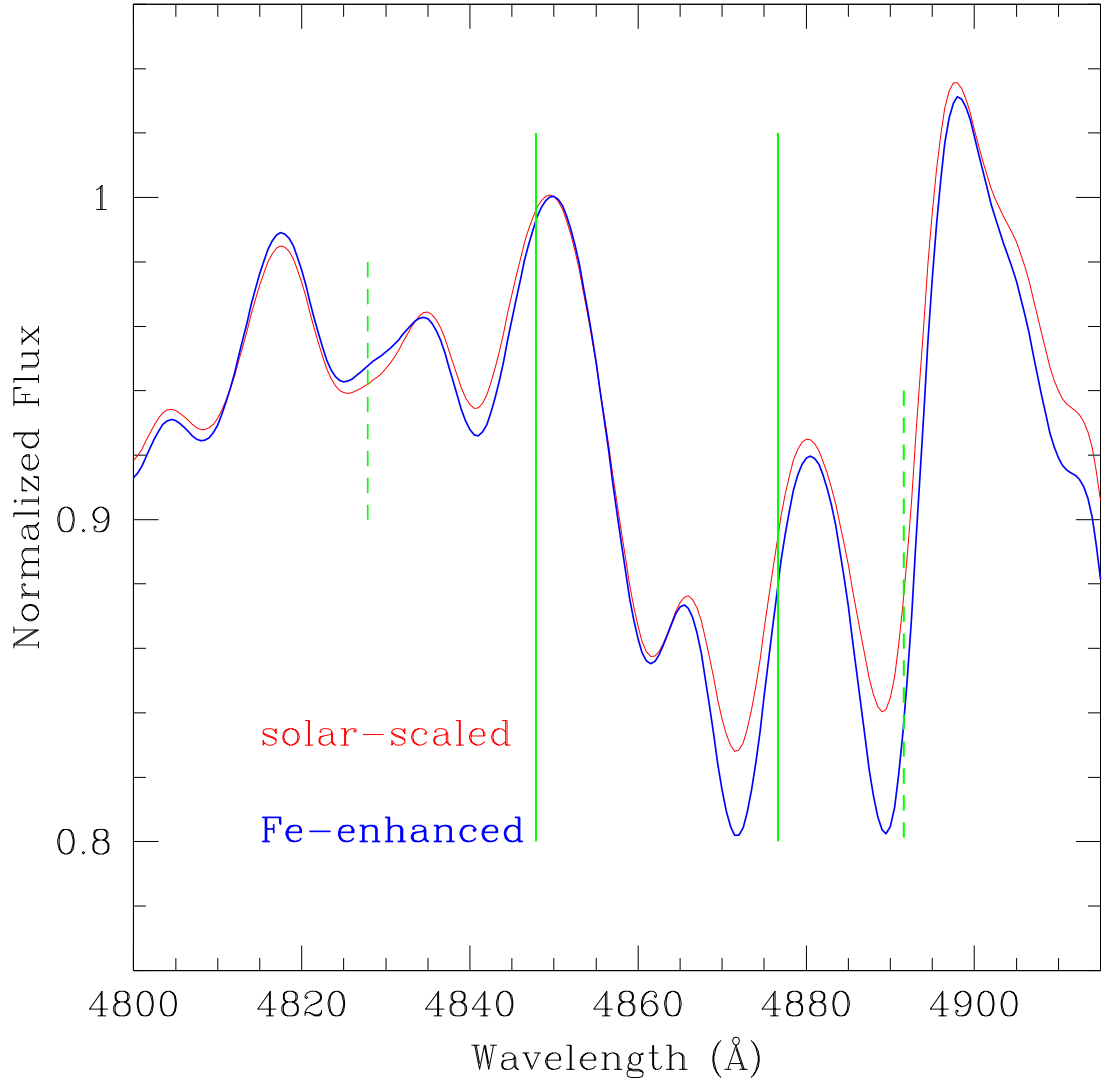}
\caption{Solar-scaled and Fe-enhanced 12 Gyr integrated spectra near 
H$\beta$ (4861 \AA).  Note that Fe-lines make the index bandpass 
and the red-continuum levels significantly lower, which consequently 
cancel out their effects on the H$\beta$ index strengths 
as we see in the lower right panel of Figure 10. \label{fig13}}
\end{figure}

\begin{figure}
\epsscale{1.}
\plotone{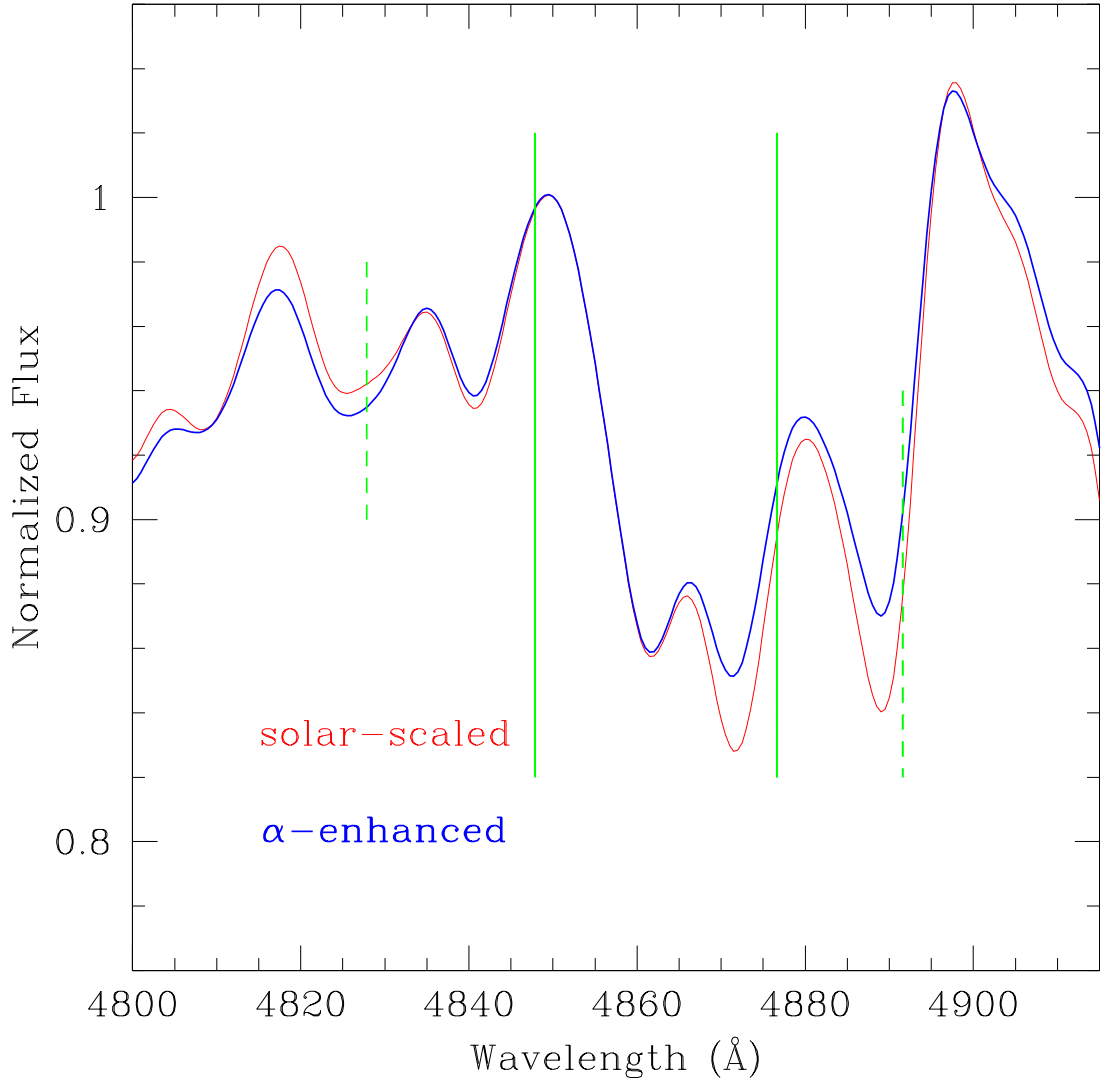}
\caption{Solar-scaled and the $\alpha$-enhanced 12 Gyr integrated spectra near 
H$\beta$ (4861 \AA).  Note that the $\alpha$-enhanced spectra at constant $Z$ 
mostly reflect the Fe depression effect which is the opposite of what is seen 
in Figure 13 and they affect the H$\beta$ index strengths little as we see 
in the lower right panel of Figure 10. \label{fig14}}
\end{figure}

\begin{figure}
\epsscale{1.}
\plotone{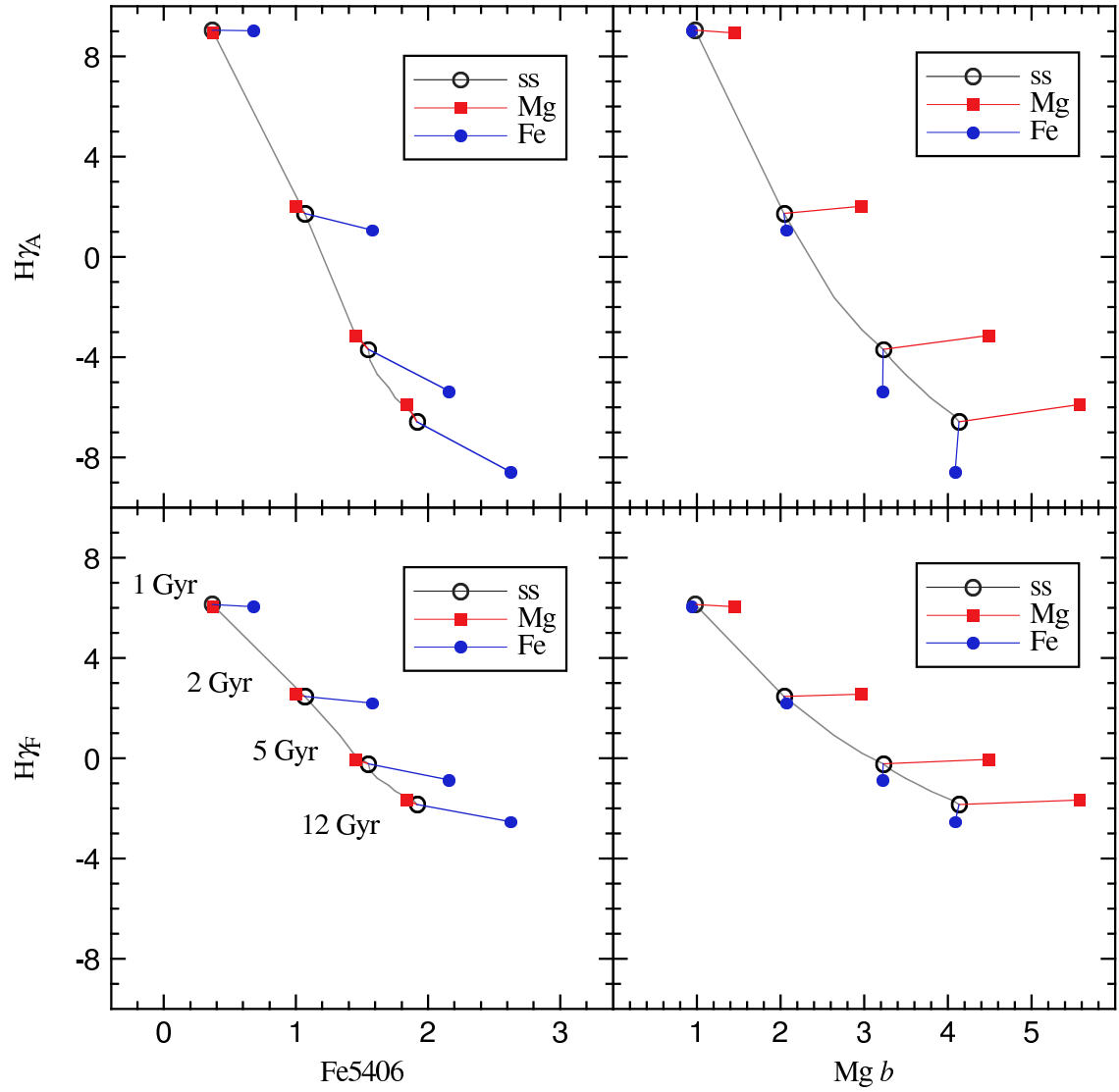}
\caption{H$\gamma_{A}$ (top) and H$\gamma_{F}$ (bottom) are compared 
with Fe5406 (left) and Mg $b$ (right).  Clearly H$\gamma_{A}$ is significantly 
affected by iron enhancement. \label{fig15}}
\end{figure}

\begin{figure}
\epsscale{1.}
\plotone{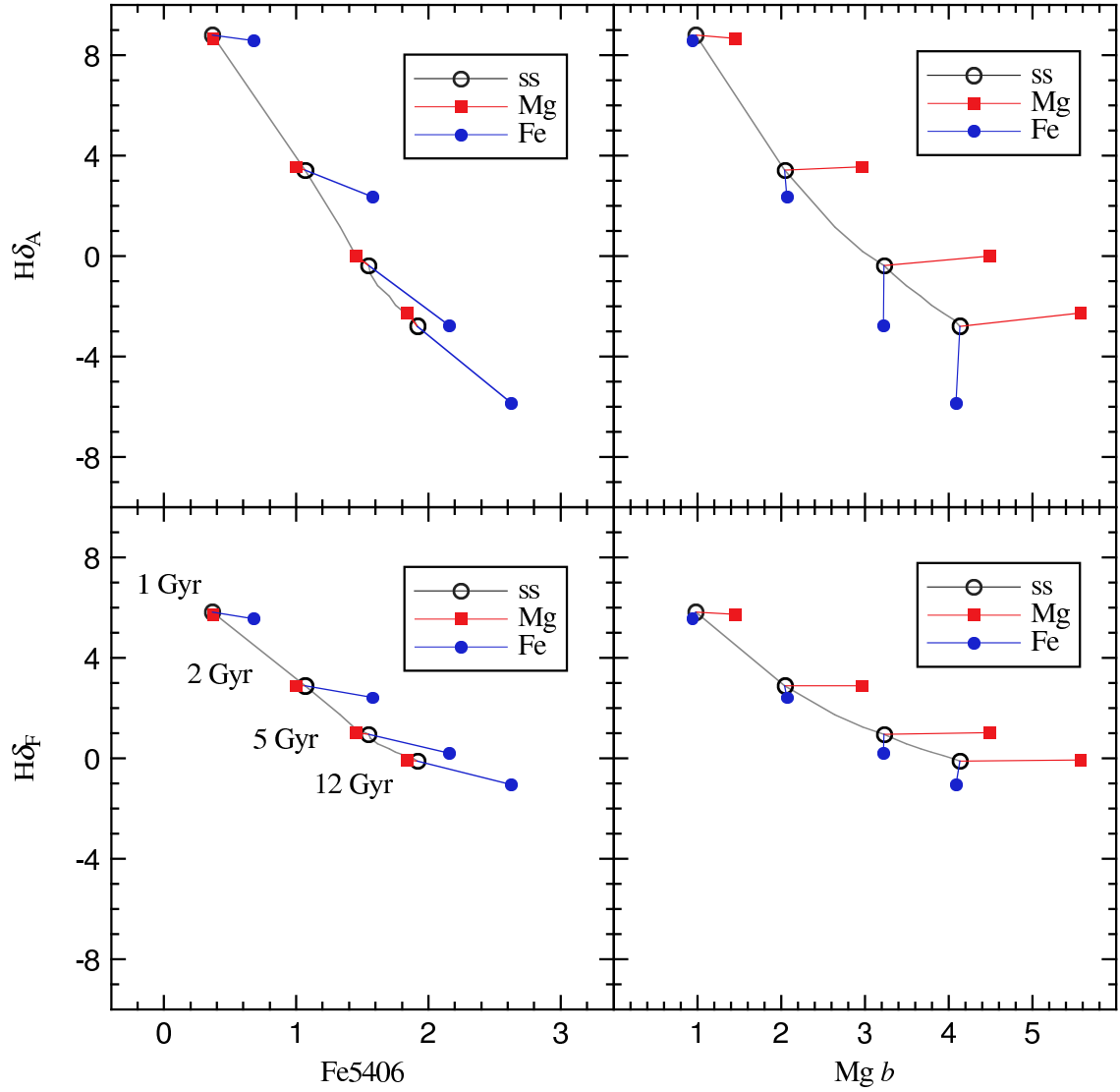}
\caption{Similar to Figure 15, but H$\delta_{A}$ (top) 
and H$\delta_{F}$ (bottom) are compared with Fe5406 (left) 
and Mg $b$ (right).  Again, clearly H$\delta_{A}$ 
is significantly affected by iron enhancement. \label{fig16}}
\end{figure}

\begin{figure}
\epsscale{1.}
\plotone{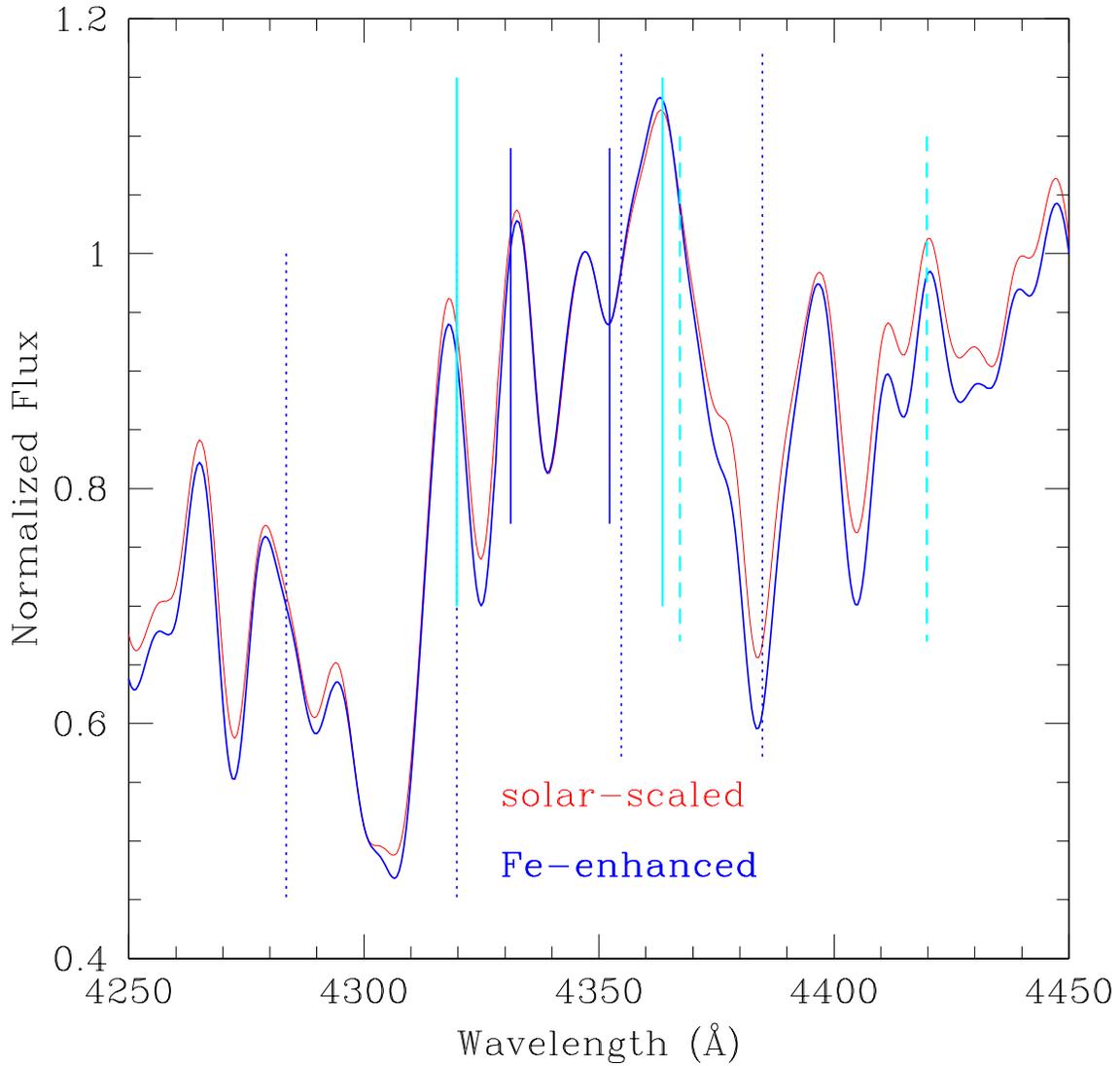}
\caption{Solar-scaled and Fe-enhanced 12 Gyr integrated 
spectra near H$\gamma$ (4341 \AA).  The solid lines and the dashed lines 
are the index bandpasses (the broader ones are the $A$ indices, 
while the narrower ones are the $F$ 
indices) and the continua regions for the $A$ indices, respectively, 
and the dotted lines are the continua regions for the $F$ indices. 
H$\gamma_{A}$ and H$\gamma_{F}$ have the same blue continuum.  
Note that the iron 
lines make the red-continuum levels significantly lower, which 
consequently make the H$\gamma$ index strengths weaker, 
especially for the broader H$\gamma_A$ index as we see 
in the upper panels of Figure 15. \label{fig17}}
\end{figure}

\begin{figure}
\epsscale{1.}
\plotone{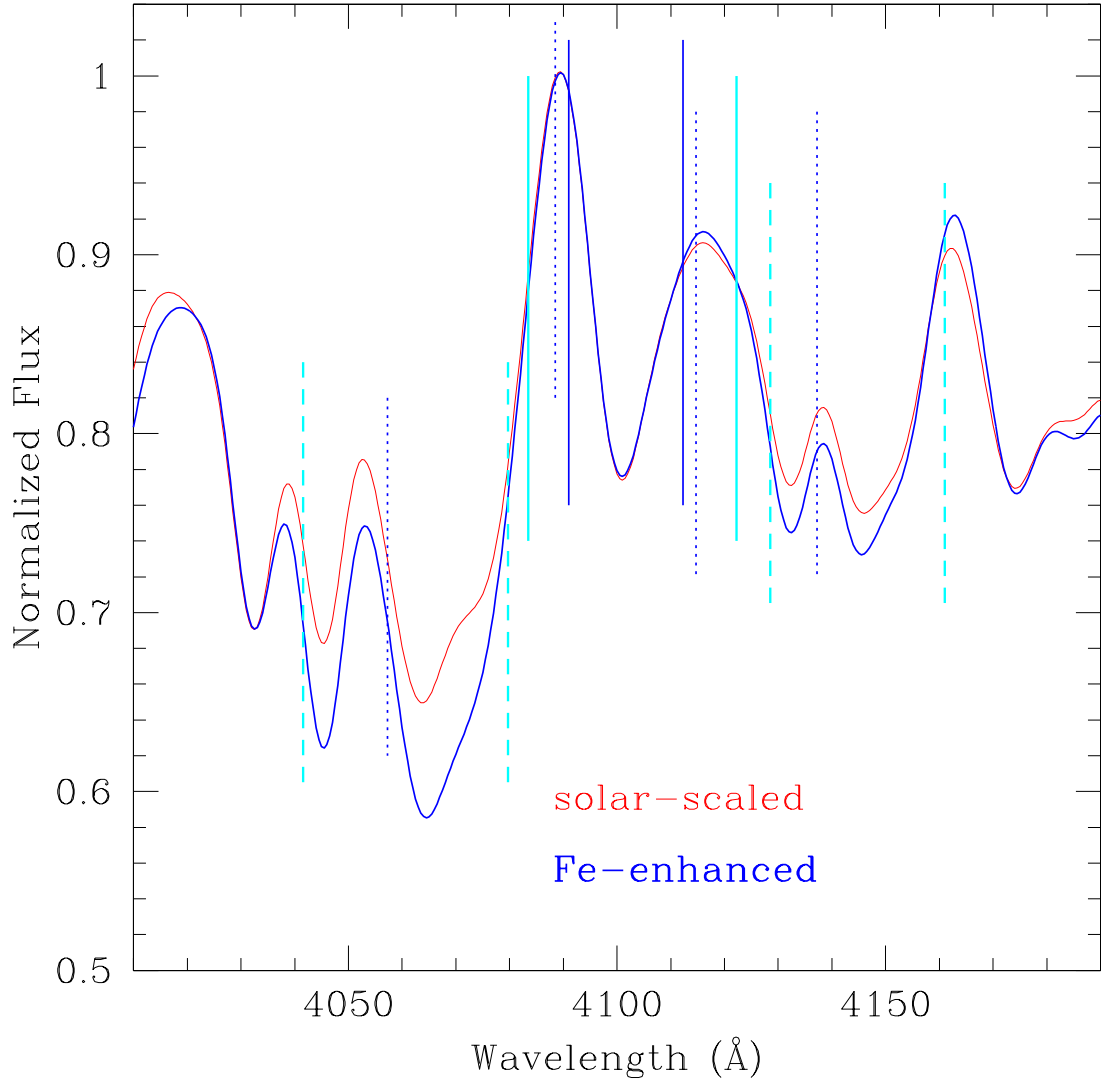}
\caption{Solar-scaled and Fe-enhanced 12 Gyr integrated 
spectra near H$\delta$ (4102 \AA).  Symbols are same as in Figure 17.  
Note that the iron 
lines make the continua levels significantly lower, which 
consequently make the H$\delta$ index strengths weaker, 
especially for the broader H$\delta_A$ index as we see 
in the upper panels of Figure 16. \label{fig18}}
\end{figure}

\clearpage

\begin{figure}
\epsscale{1.}
\plotone{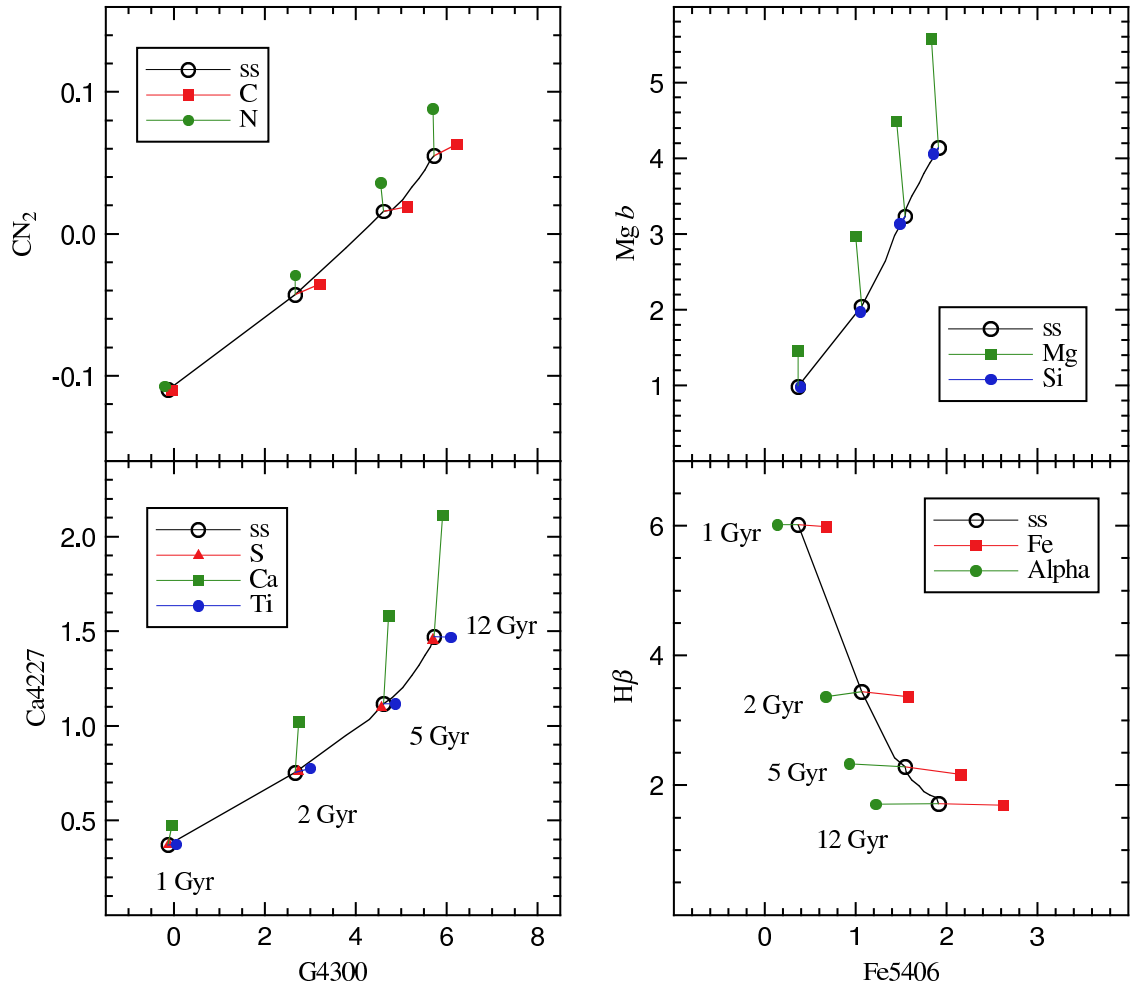}
\caption{The most conspicuously element-sensitive cases are collectively 
displayed here.  Left: G4300 vs. CN$_2$ (top) and Ca4227 (bottom).  
Right: Fe5406 vs. Mg $b$ (top) and H$\beta$ (bottom). \label{fig19}}
\end{figure}

\begin{figure}
\epsscale{1.}
\plotone{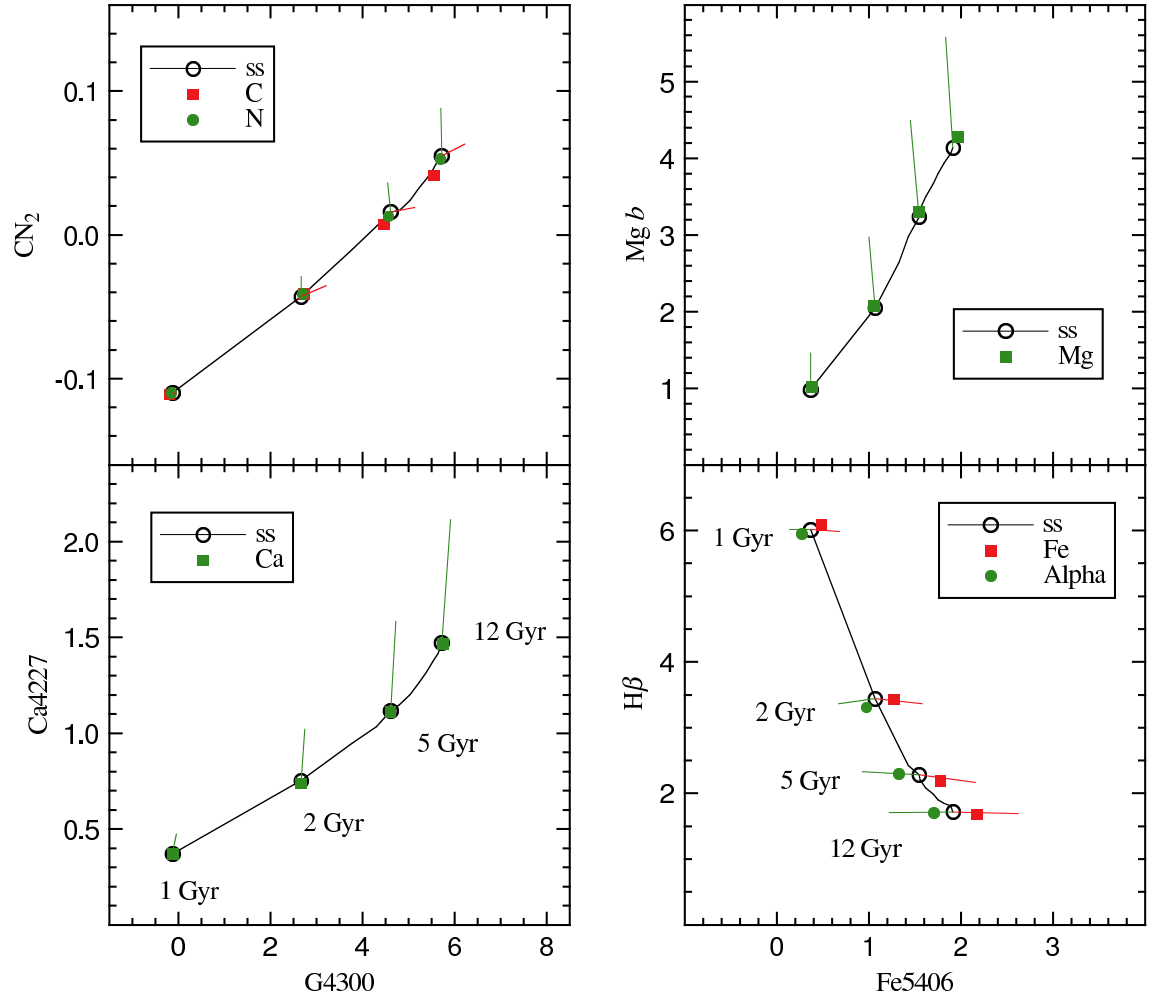}
\caption{Similar to Figure 19, but now filled symbols denote isochrone 
effects alone, while the lines show the combination of 
isochrone and stellar spectral effects.  It is clear from 
this figure that stellar spectral effects dominate the changes in the 
indices, whereas the isochrone effects are comparatively 
insignificant (see text). \label{fig20}}
\end{figure}

\clearpage

\begin{figure}
\epsscale{1.}
\plotone{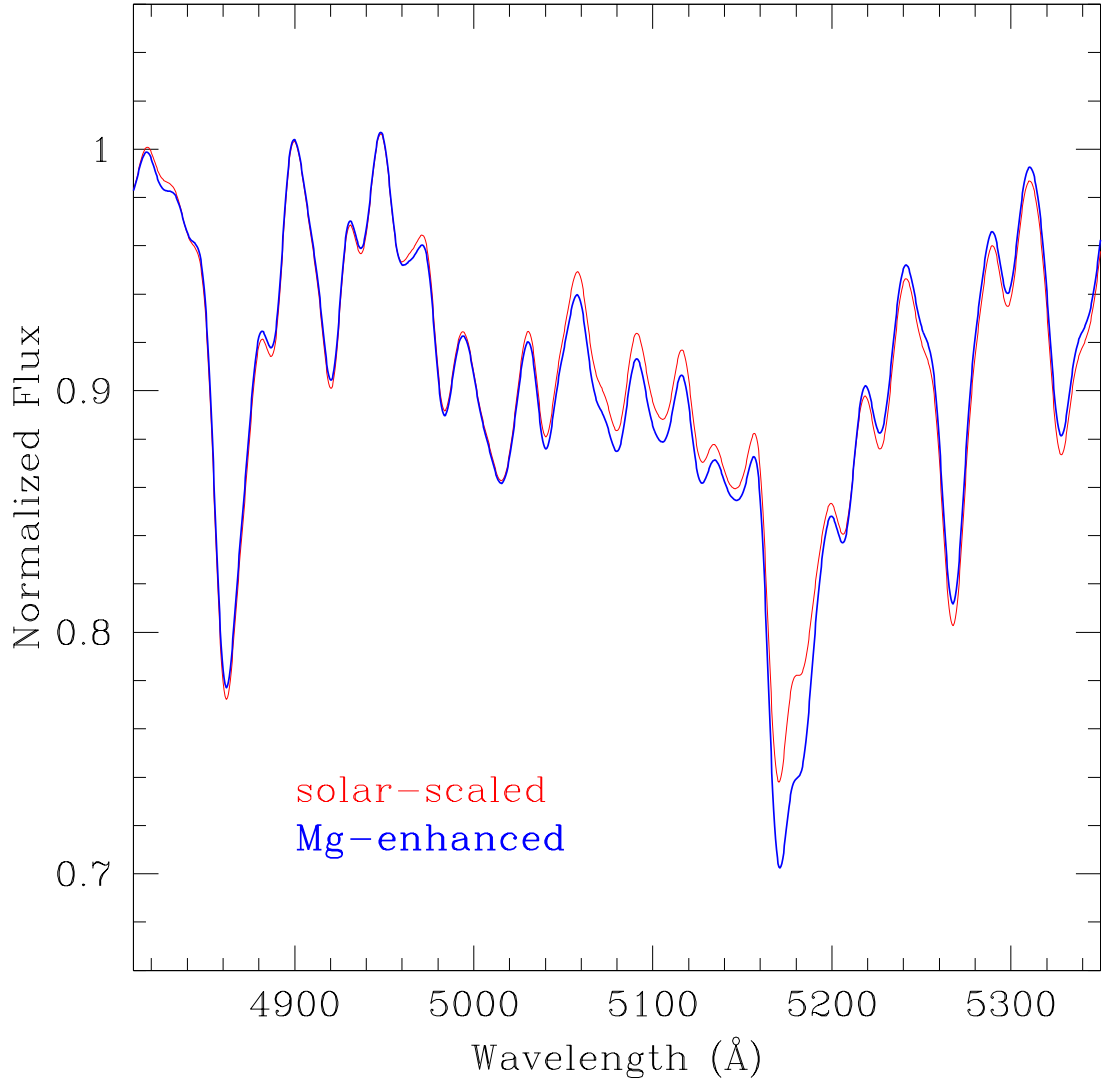}
\caption{2 Gyr-old, scaled-solar spectrum smoothed to $300\,km\,s^{-1}$ 
compared with a 2 Gyr-old, +0.3 dex magnesium-enhanced spectrum with 
the same $Z$ in the SAURON spectral range.  The spectra are normalized at 
4750 \AA.  Note that Mg$_1$ ($\sim$ 5100 \AA) 
and Mg $b$ ($\sim$ 5175 \AA) become stronger with magnesium 
enhancement. \label{fig21}}
\end{figure}

\begin{figure}
\epsscale{1.}
\plotone{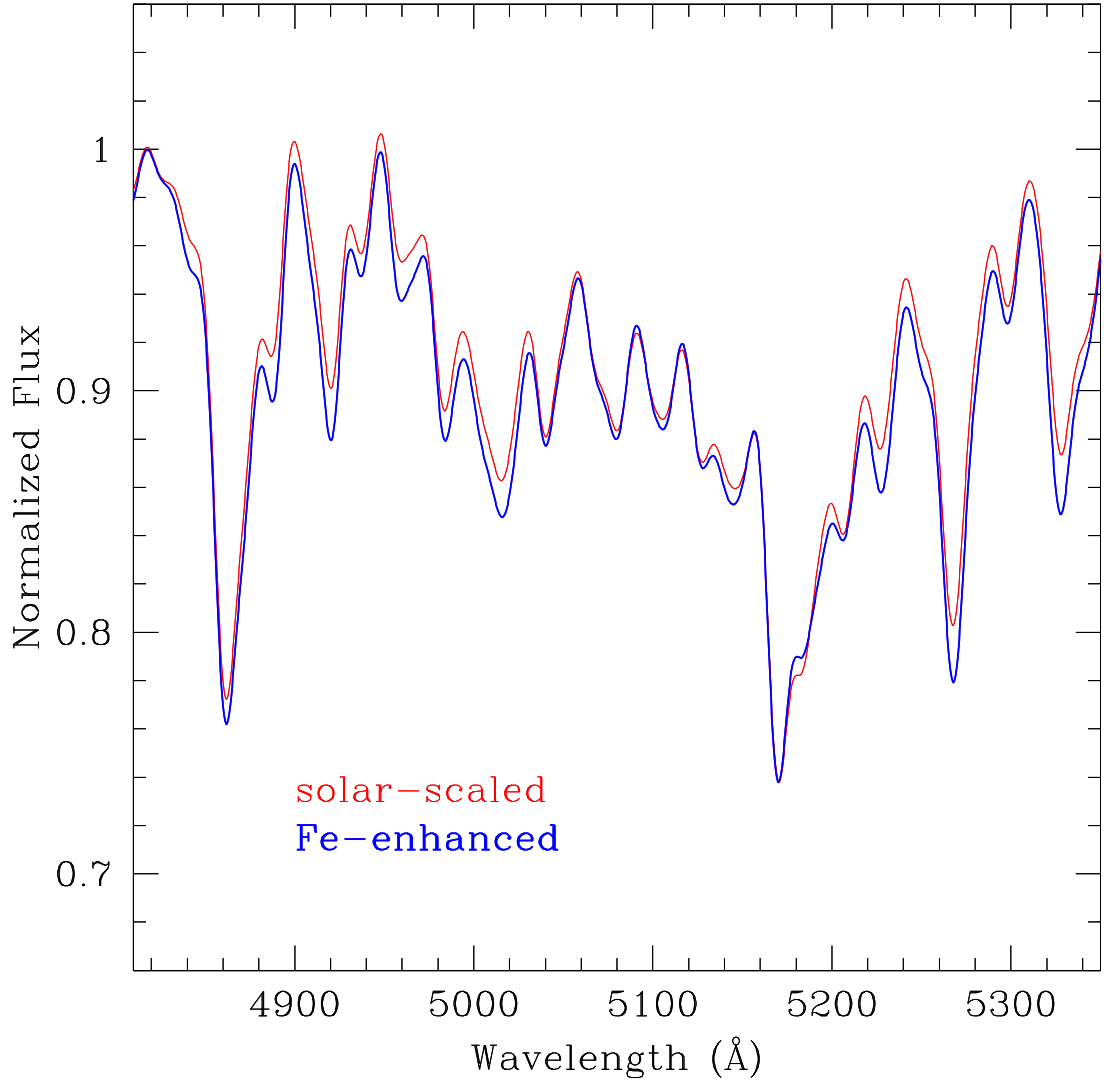}
\caption{2 Gyr-old, scaled-solar spectrum smoothed to $300\,km\,s^{-1}$ 
compared with a 2 Gyr-old, +0.3 dex iron-enhanced spectrum with 
the same $Z$ in the SAURON spectral range.  The spectra are normalized at 
4750 \AA.  Note that Fe5015, 
Fe5270, and Fe5335 become stronger with iron enhancement. \label{fig22}}
\end{figure}

\clearpage

\begin{figure}
\epsscale{1.}
\plotone{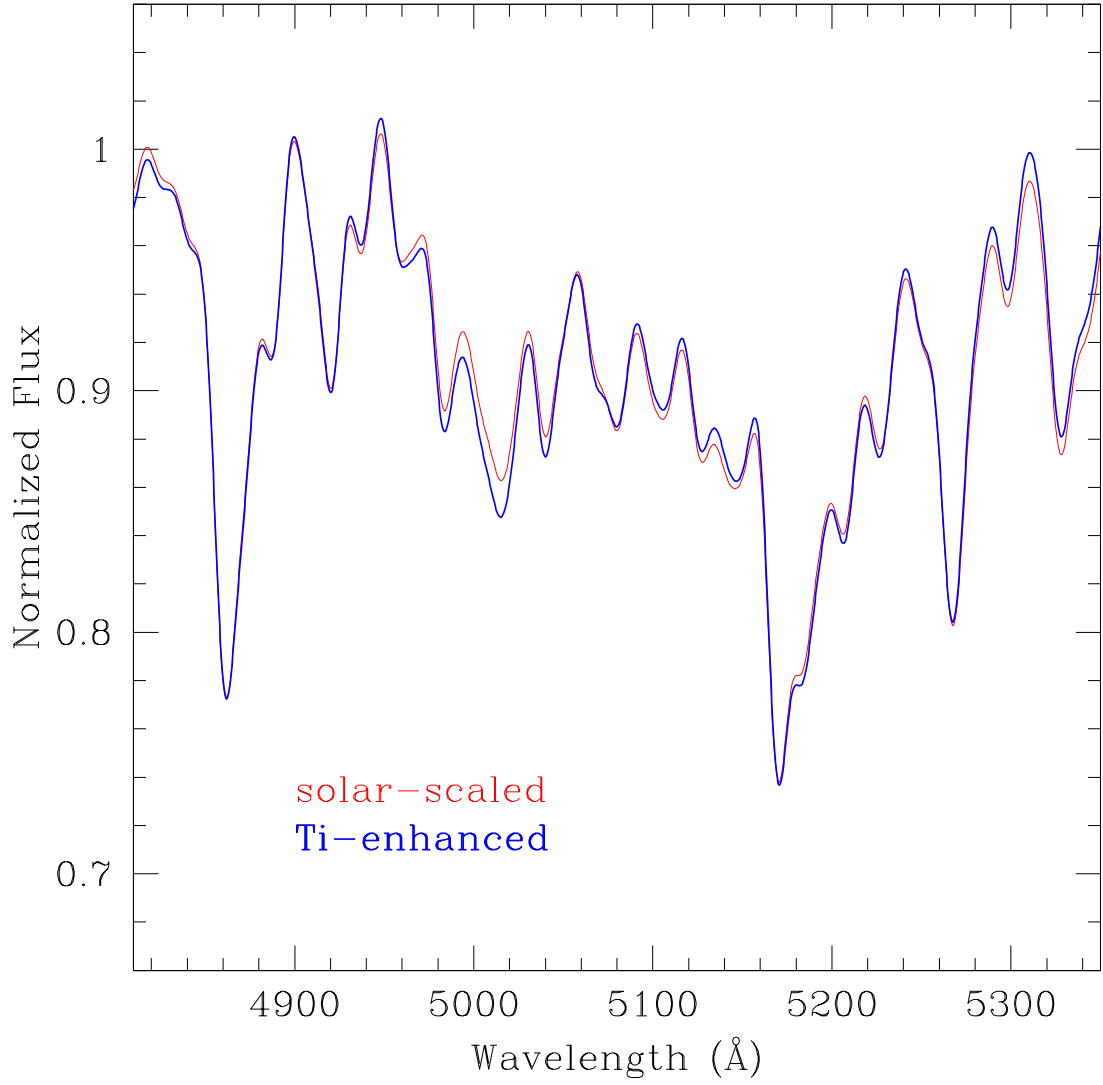}
\caption{2 Gyr-old, scaled-solar spectrum smoothed to $300\,km\,s^{-1}$ 
compared with a 2 Gyr-old, +0.3 dex titanium-enhanced spectrum with 
the same $Z$ in the SAURON spectral range.  The spectra are normalized at 
4750 \AA.  Note that Fe5015 becomes stronger with titanium enhancement, 
while the rest of the indices are mostly intact. \label{fig23}}
\end{figure}

\begin{figure}
\epsscale{1.}
\plotone{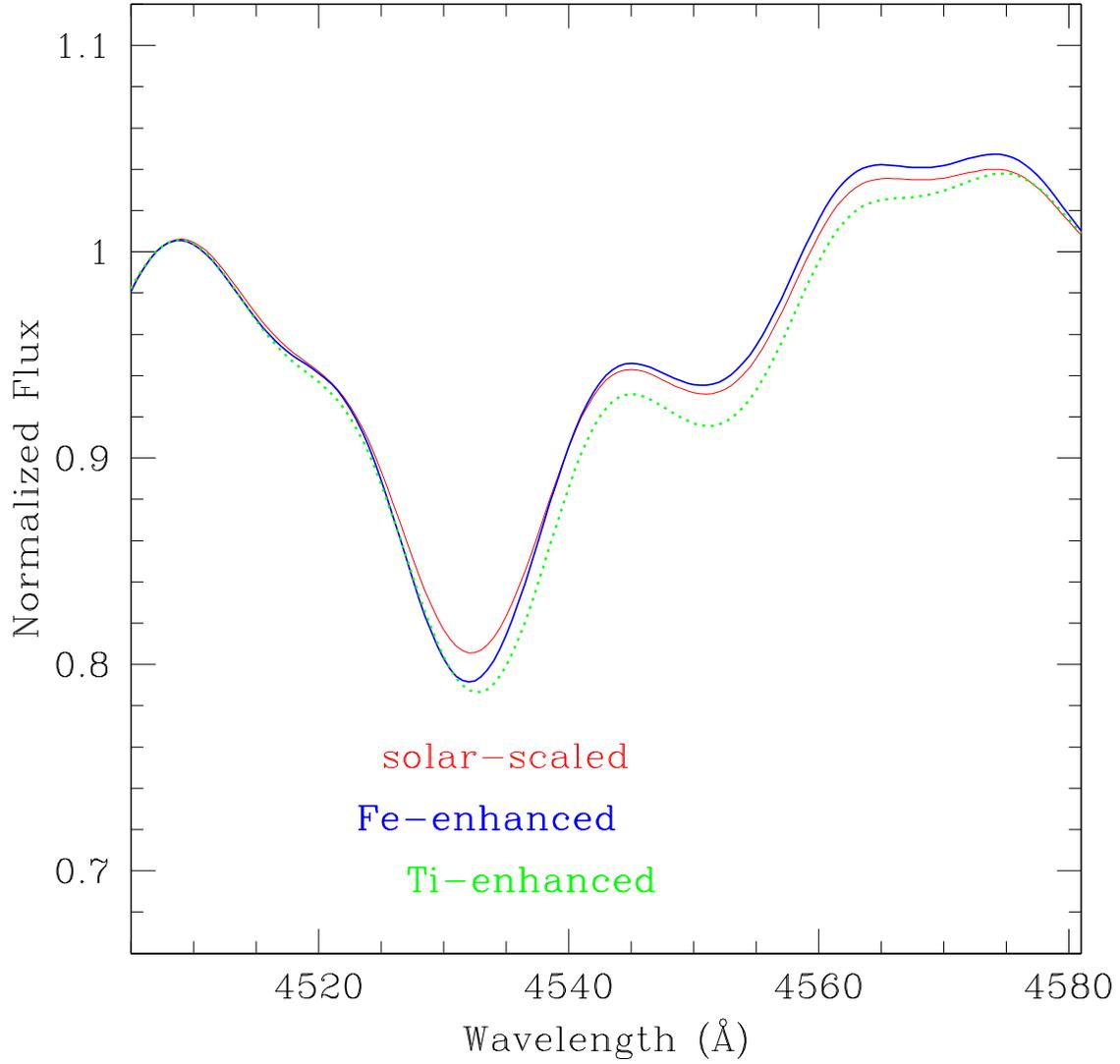}
\caption{Solar-scaled, Fe-enhanced, and Ti-enhanced 12 Gyr 
integrated spectra in the Fe4531 index region.  Note that the 
centroids of Fe-enhanced and Ti-enhanced (dotted line) spectra are located 
to the left and right side of the solar-scaled one, 
respectively, because of different locations of Fe and Ti lines within 
the Fe4531 index bandpass. \label{fig24}}
\end{figure}

\end{document}